\documentclass[usenatbib]{mn2e}
\usepackage{graphicx}
\newcommand{\Fs}{\,^*\! F}

\newcommand{\bm}{\bmath{m}}

\newcommand{\bB}{\bmath{B}}
\newcommand{\bE}{\bmath{E}}
\newcommand{\bH}{\bmath{H}}
\newcommand{\bD}{\bmath{D}}
\newcommand{\bL}{\bmath{L}}

\newcommand{\bS}{\bmath{S}}

\newcommand{\text}[1]{\quad\mbox{#1}\quad}
\newcommand{\spr}[2]{\bmath{#1} \!\cdot\! \bmath{#2}}
\newcommand{\vpr}[2]{\bmath{#1} \!\times\! \bmath{#2}}

\newcommand{\sub}[1]{_{\mbox{\tiny #1}}}

\title[Activation of the BZ mechanism]
{Activation of the Blandford-Znajek mechanism in collapsing stars} 

\author[S.S. Komissarov \&  M.V.Barkov]{
Serguei S.~Komissarov,$^{1}$
Maxim V.~Barkov,$^{1,2}$\thanks{ 
email: serguei@maths.leeds.ac.uk, bmv@maths.leeds.ac.uk}
\\$^{1}$Department of Applied Mathematics, The University of Leeds,
Leeds, LS2 9GT\\
$^{2}$Space Research Institute, 84/32 Profsoyuznaya Street, Moscow
117997, Russia}
\begin{document}
\date{Received/Accepted}
\maketitle

\begin{abstract}
Collapse of massive stars may result in formation of accreting black holes 
in their  interior. The accreting stellar matter may advect substantial magnetic 
flux onto the black hole and promote release of its rotational energy via 
magnetic stresses (the Blandford-Znajek mechanism).  In this paper we explore 
whether this process can explain the stellar explosions and relativistic jets 
associated with long Gamma-ray-bursts. In particularly, we show that the 
Blandford-Znajek mechanism is activated when the rest mass-energy density of 
matter drops below the energy density of magnetic field in the very vicinity 
of the black hole (within its ergosphere). We also discuss whether such a strong 
magnetic field is in conflict with the rapid rotation of stellar core required in 
the collapsar model and suggest that the conflict can be avoided if the 
progenitor star is a component of close binary. In this case the stellar 
rotation can be sustained via spin-orbital interaction. In an alternative 
scenario the magnetic field is generated in the accretion disk 
but in this case the magnetic flux through the black hole ergosphere is not
expected to be sufficiently high to explain the energetics of hypernovae by 
the BZ mechanism alone. However, this energy deficit can be recovered via 
additional power provided by the disk.  

\end{abstract}
                                                                                          
                                                                                          
\section{Introduction}
\label{introduction}

The most popular model for the central engines of long Gamma-ray-burst (GRB) 
jets is based on the ``failed supernova'' scenario of stellar collapse, 
or ``collapsar'', where the iron core of progenitor star forms a BH \citep{W93}.  
If the specific angular momentum in the equatorial part of progenitor exceeds 
that of the marginally bound orbit of the BH then the collapse becomes highly 
anisotropic.  While in the polar
region it may proceed more or less uninhibited, at least for a while, the equatorial
layers form dense and massive accretion disk. 
The gravitational energy released in this disk can be very large, more then
sufficient to overturn the collapse of outer layers  and drive GRB outflows 
\citep{MW99}.  A similar configuration can be produced via inspiraling of a BH 
or a neutron star into the companion star during the common envelope phase of 
a close binary \citep{TI79,FW98,ZF01}. In any case, the observed duration of long 
GRBs, between 2 and 1000 seconds, imposes a strong constraint on the size of 
progenitors - their light crossing time should be significantly shorter then the 
burst duration. This implies that the progenitors must be compact stars stripped of 
their hydrogen envelopes.  This conclusion agrees with the absence of hydrogen 
lines in the spectra of supernovae identified with GRBs (Moreover, since some of
these supernovae are of Type Ic their progenitors may have lost their 
helium envelopes as well.).  
Massive solitary stars can lose their envelopes via 
strong stellar winds.  However, this may lead to unacceptably low rotation rates 
by the time of stellar collapse \citep{HWS05}. In the case of binary system 
the loss of envelope can be caused by the gravitational interaction with companion.  

The currently most popular mechanism of powering GRB jets is the heating via 
annihilation of neutrinos and anti-neutrinos produced in the disk \citep{MW99}. 
The energy deposited in this way is a strong function of the
mass accretion rate which must be above $\simeq 0.05 M_\odot\mbox{s}^{-1}$ in
order to agree with the observational constraints on the energetics of GRBs
\citep{PWF99}. Such high accretion rates can be provided in the collapsar
scenario and the numerical simulations by \citet{MW99} and \citet{AIMGM00} 
have demonstrated that sufficiently large energy deposition in the polar region 
above the disk may indeed result in fast collimated jets.  
However, we have to wait for simulations with proper implementation of neutrino 
transport before making final conclusion on the suitability of this model -- 
the long and complicated history of numerical studies of neutrino-driven supernova 
explosions teaches us to be cautious. The first attempt to include neutrino transport 
in collapsar simulations was made by \citet{NTMT07} and they did not see 
neutrino-driven polar jets - the heating due to neutrino annihilation could not 
overcome the cooling due to neutrino emission.\footnote{However, they have
simulated only the very initial stages of the collapsar evolution, $t<2.2$s,
whereas the jet simulation of \citet{MW99} started at
$t=7$s when the density in the polar regions above the black hole has dropped
down to $10^6\mbox{g}\,\mbox{cm}^{-3}$ leading to much lower cooling
rates.} 
In any case, it is 
unlikely that this model can explain the bursts that are longer than
$100$s as by this time the mass accretion rate is expected to drop significantly. 
For example the accretion rate in the simulations by 
\citet{MW99} has already drifted below the critical $\dot{M}\simeq
0.05 M_\odot\mbox{s}^{-1}$ by $t=20$s.
 
The two alternatives to the neutrino mechanism which are also frequently 
mentioned in connection with GRBs are the magnetic braking of the accretion 
disk \citep{BP82,NPP92,MR97,PMAB03,UM06} and the magnetic braking of the central black 
hole \citep{BZ77,MR97,LBW00,M06,BK08,BK08b}. 

A number of groups have studied the potential of magnetic mechanism in the
collapsar scenario using Newtonian MHD codes and implementing the
Paczynski-Witta potential in order to approximate the gravitational field of
central BH \citep{PMAB03,FKYHS06,NTMT07}. In this approach it is impossible to
capture the Blandford-Znajek effect and only the magnetic braking of 
accretion disk can be investigated. The general conclusion of these studies is
that the accretion disk can launch magnetically-driven jets provided the
magnetic field in the progenitor core is sufficiently strong. This field is
further amplified in the disk, partly due to simple winding of poloidal
component and partly due to the magneto-rotational instability (MRI), until
the magnetic pressure becomes very large and pushes out the surface layers of
the disk. In particular, \citet{PMAB03} studied the collapse of a star with a
similar structure to that considered  \citet{MW99} and with purely monopole 
initial magnetic field of strength $B=2\times10^{14}$G at $r=3r_g$. The corresponding 
total magnetic flux, $\Psi_h\simeq 10^{26}\mbox{G}\,\mbox{cm}^2$, is in fact within
the observational constraints on the field of magnetic stars, including white
dwarfs \citep{S03}, Ap-stars \citep{FW05} and massive O-stars \cite{D02}.  They
used realistic equation of state and included the neutrino cooling but not 
the heating.  The simulations lasted up to $t\simeq
0.28$s in physical time during which a Poynting-dominated polar jet has
developed in the solution. 
The jet power seemed to show strong systematic
decline, from $\simeq 10^{51}\mbox{erg}\,\mbox{s}^{-1}$ down to $\simeq
10^{50}\mbox{erg}\,\mbox{s}^{-1}$ or even less, which might signal
transient phenomenon.

\citet{SS07} studied the collapse of rotating stellar
cores and formation of BHs and their disks in the collapsar scenario 
using the full GRMHD approximation. Their simulations show
powerful explosions soon after the accretion disk is formed and
the free falling plasma of collapsing star begins to collide with this disk. 
However, they did not account for the neutrino cooling
and the energy losses due to photo-dissociation of atomic nuclei so these
explosions could be similar in nature to the ``successful'' prompt explosions
of early supernova simulations \citep{Bethe}. 
\citet{MYKS04a,MYKS04b} carried out GRMHD simulations in the
time-independent space-time of a central BH.  Their computational domain did not
capture the BH ergosphere and thus they could not study the role of the
Blandford-Znajek effect~\citep{K04a,K08}. In addition, the energy gains/losses 
due to the neutrino heating/cooling were not included
and the equation of state (EOS) was a simple polytrope. The simulations 
run for a rather short time, $\simeq 280 r_g/c$ where $r_g=GM/c^2$, and jets
were formed almost immediately, presumably due to the extremely strong initial 
magnetic field. 

The original theory of the Blandford-Znajek mechanism was developed 
within the framework of magnetodynamics (the singular limit of RMHD 
approximation where the inertia of plasma particles is fully ignored). 
The pioneering numerical studies of this mechanism have shown that it 
can also operate within the full RMHD approximation, at least in 
the regime where the energy density of matter is much smaller compared to 
that of the electromagnetic field, and that it can be captured by modern 
numerical techniques \citep{K04b,Koi04,MG04,N09}. However, no systematic 
numerical study has been carried out in order to establish the exact 
conditions for activation of the BZ mechanism.        

The first simulations demonstrating the potential of the BZ mechanism in the 
collapsar scenario were carried out by \citet{M06}
who also used simple polytropic EOS and considered purely adiabatic case, but
captured the effects of black hole ergosphere. The initial configuration
included a torus with poloidal magnetic and a free-falling stellar envelope. The
results show ultrarelativistic jets with the Lorentz factor up to 10 emerging
from the black hole indicating that the Blandford-Znajek effect may play a key
role in the production of GRB jets. \citet{BK08,BK08b} 
added a realistic EOS and included the energy losses due to neutrino emission 
(assuming optically thin regime) and photo-dissociation of nuclei. They 
also reported the development of relativistic outflows powered by the 
central black hole via the Blandford-Znajek mechanism (the  
magnetic braking of accretion disk did not seem to contribute much to the 
jet power though).      

The black hole-driven magnetic stellar explosions and relativistic 
jets observed in these exploratory computer simulations invite a deeper
analysis of this  model. Obviously, such an outcome cannot be a generic 
feature of core-collapse supernovae. Indeed, only a very small fraction of 
SNe~Ib/Ic seem to produce GRBs \citep{P05,WB06}. So what are the relevant 
conditions and can they be reproduced as the results of stellar evolution?  
These are the main issues we address in this paper.

\section{Blandford-Znajek mechanism and GRBs}     
\label{BZ-theory}

The rotational energy of Kerr black hole is
\begin{equation} E\sub{rot} = M_hc^2 f_1(a) \,\simeq\, 1.8\!\times\! 10^{54}
f_1(a) \left(\frac{M_h}{M_\odot} \right) \mbox{erg},
\label{e-rot}
\end{equation}
where
$$
f_1(a) = 1-\frac{1}{2}\left[ \left(1+\sqrt{1-a^2}\right)^2+a^2 \right]^{1/2},
$$
$M_h$ is the BH mass and $a\in[0,1)$ is its dimensionless rotation
parameter. For $M\sub{bh}=2M_{\sun}$ and $a=0.9$ this gives the enormous value
of $E\sub{rot} \simeq 5\times10^{53}$erg which is about fifty times higher
than the rotational energy of a millisecond pulsar and well above of what is
needed to explain the energy of GRBs and associated hypernovae. Even for a
relatively slowly rotating black hole with $a=0.1$ this is still a respectable
number,  $2.3\!\times\!10^{51}$erg.  Moreover, because in all versions of 
the collapsar model the black hole spin is aligned with the spin of accretion disk
this energy reserver is continuously replenished via accretion. 
Thus, as far as the availability of energy is concerned the black hole model 
looks very promising indeed.

The energy release rate is usually estimated using the 
Blandford-Znajek power for the case of force-free monopole 
magnetosphere 
\begin{equation}
\dot{E}\sub{BZ}=\frac{1}{6c}\left(\frac{\Omega_h\Psi_h}{8\pi}\right)^2,
\label{e-bz0}
\end{equation} 
where $\Omega_h$ is the angular velocity of the BH and $\Psi_h$
is the magnetic flux threading one hemisphere of the BH horizon (Here it is
assumed that the angular velocity of the BH magnetosphere 
$\Omega=0.5\Omega_h$.). This formula is quite accurate not only for slowly 
rotating black holes considered in \citet{BZ77} but also for 
rapidly rotating BHs \citep{K01}.  In application to the collapsar problem it gives 
us the following estimate
\begin{equation} \dot{E}\sub{BZ}=1.4\times10^{51} f_2(a) \Psi\sub{h,27}^2
\left(\frac{M_h}{M_\odot}\right)^{-2} \,\mbox{erg}\, \mbox{s}^{-1},
\label{e-bz}
\end{equation} 
where 
$$
f_2(a)=a^2\left(1+\sqrt{1-a^2}\right)^{-2},
$$ 
and $\Psi\sub{h,27}=\Psi_h/10^{27}\mbox{G}\,\mbox{cm}^2$.  One can see that
the power of BZ-mechanism is rather sensitive to black hole's mass and
magnetic flux. Since, the black hole mass is likely to be $\ge 3M_\odot$ 
the observed energetics of hypernovae and long GRBs requires 
$\Psi\sub{h,27}\simeq 1$. This value is comparable with
the maximal surface flux observed in magnetic stars, Ap stars, 
magnetic white dwarfs, and magnetars \citep[e.g.][]{FW05}. Thus, the
magnetic field of GRB central engine may well be the original field of the
progenitor star.

The estimates (\ref{e-rot},\ref{e-bz}) show that braking of black holes alone
can explain the energetics of GRBs and this is why this mechanism is often
mentioned in the literature on GRBs. However, there is another issue to take
into consideration. The equation (\ref{e-bz0}) is obtained in the limit where
the inertia of magnetospheric plasma and, to large degree, its gravitational 
attraction towards the black hole are ignored. In contrast, the mass density of 
plasma in the collapsing star may be rather high and has to be taken into
account. For example, the magnetohydrodynamic waves may become trapped 
in the accretion flow and unable to propagate outwards. In such a case, it 
would not be possible to extract magnetically any of the black hole rotational 
energy and drive stellar explosion irrespectively of how high this energy 
is.  Since the BZ mechanism can only operate within the black hole ergosphere 
\citep[e.g.][]{K04a,K08}, the magnetohydrodynamic waves must at least be able to 
cross the ergosphere in the outward direction for its activation. 
This agrees with the conclusion reached in \citet{T90} that the Alfven surface 
of a steady-state ingoing wind must be located inside the ergosphere to have 
outgoing direction of the total energy flux.          
Thus we should consider the following condition for activation of the 
BZ-mechanism in accretion flows: the Alfven speed has to exceed
the local free fall speed at the ergosphere,
$$
  v_a>v_{f}.
$$
In order to simplify calculations let us apply the Newtonian expressions for
both speeds. The Alfven speed is
$$
   v_a^2 = \frac{B^2}{4\pi\rho},
$$
where $B$ is the magnetic field strength and $\rho$ is the plasma mass
density, whereas the free-fall speed is
$$
   v^2_{f}=\frac{2GM_h}{r}.
$$ 
At $r=2r_g=2GM/c^2$ the local criticality condition reads
\begin{equation} \beta_\rho = \frac{4\pi\rho c^2}{B^2} < 1.
\label{cond-1}
\end{equation} That is the energy density of magnetic field has to exceed that
of plasma in the vicinity the black hole.

For spherically symmetric flows the condition $v_a>v_{f}$ can be written as a
constraint on the mass accretion rate, $\dot{M}$, and the flux of radial 
magnetic field, $\Psi$,
$$
\frac{\Psi}{2\pi r \sqrt{\dot{M}v_{f}(r)}} >1.
$$
Since $v_{f}\propto r^{-1/2}$ this condition is bound not to be satisfied at
large $r$ but we only need to apply it at the ergosphere.  Using $r\simeq
2r_g$ we then obtain
\begin{equation} \kappa > \kappa_c,
\label{cond-2}
\end{equation} where
\begin{equation} \kappa = \frac{\Psi_h}{4\pi r_g \sqrt{\dot{M}c}},
\end{equation} and $\kappa_c=1$. In fact, the critical value of $\kappa$ must
depend on the black hole spin and tend to infinity for small $a$. The
Newtonian analysis cannot capture this effect. The relativistic one, on the
other hand, appears rather complicated \citep{C89,T90}. Fortunately, nowadays 
this issue can be investigated via numerical simulations. Another, interesting 
and important issue for time-dependent simulations is whether there is only 
one bifurcation separating solutions with switched-on and switched-off BZ 
mechanism or the transition is reacher and allows more complicated time-dependent 
solutions, e.g. quasi-periodic or chaotic ones.

\section{Test simulations.}
\label{test-sim}

The first type of simulations presented in this paper are designed to test the
validity of our arguments behind the criterion for activation of
BZ-mechanism which we derived in Sec.\ref{BZ-theory} using a combination of
Newtonian and relativistic physics.  For this purpose we consider a more or
less spherical accretion of cold magnetized plasma with vanishing angular
momentum onto a rotating black hole. In order to focus on the MHD aspects of
the problem we ignore the microphysics important in the collapsar problem and
consider plasma with simplified polytropic EOS.

The main details of our numerical method and various test simulations are
described in \citet{K99,K04b,K06}. The only really new feature here is the
introduction of HLL-solver which is activated when our linear Riemann
solver fails. This usually occurs in regions of relativistically high
magnetization where the magnetic energy density exceeds that of matter by more
than one order of magnitude or in strong rarefactions.  This makes the scheme
a little bit more robust though we are still forced to use a ``density
floor'', that is a lower limit on the value of mass-energy density of 
matter\footnote{We
have found that a total switch to the HLL-solver noticeably degrades numerical
solutions via increasing numerical diffusion and dissipation. This is
particularly noticeable for slowly evolving flow components, like accretion
disks (see also \citet{MB06}.}.

Both in the numerical scheme and throughout this section of the paper 
we utilize geometric units where $G=c=M_h=1$. 
The Kerr spacetime is described using
the Kerr-Schild coordinates so the metric form reads

\begin{equation}
   \begin{array}{rl} ds^2= & g_{tt}dt^2+2g_{t\phi}dtd\phi+2g_{tr}dtdr+\\ &
g_{\phi\phi}d\phi^2 + 2g_{r\phi}d\phi dr + g_{rr}dr^2 +
g_{\theta\theta}d\theta^2,
   \end{array}
\label{KSM}
\end{equation} where
$$
  g_{tt} = \zeta-1, \quad g_{t\phi}= -\zeta a\sin^2\!\theta, \quad g_{tr} =
\zeta,
$$
$$
  g_{\phi\phi}= \Sigma \sin^2\theta/A,\quad g_{r\phi}=-a\sin^2\theta(1+\zeta),
\quad
$$
$$
  g_{rr}= 1+\zeta, \quad g_{\theta\theta}=A,
$$
and
\begin{eqnarray} \nonumber A &=& r^2 +a^2\cos^2\!\theta, \\ \nonumber \zeta
&=& 2r/A,\\ \nonumber \Sigma &=& (r^2+a^2)^2-a^2\Delta\sin^2\!\theta, \\
\nonumber \Delta &=& r^2+a^2-2r.
\end{eqnarray}

\subsection{Setup}
\label{setup-1} 

The initial solution has to describe both the flow and the magnetic field.  In
this case the flow is given by the steady-state solution for unmagnetised
cold plasma (dust). At infinity it is spherically symmetric and has vanishing
specific angular momentum and radial velocity.  The components of 4-velocity
of this flow in the coordinate basis of Kerr-Schild coordinates,
$\{\partial_\nu\}$, are
\begin{equation}
\begin{array}{l} u^t=1+\zeta\eta/(1+\eta);\\ u^\phi=-a/(A(1+\eta));\\
u^r=-\zeta \eta;\\ u^\theta=0,
\end{array}
\end{equation}
where $\eta=\sqrt{(r^2+a^2)/2r}$ and its rest mass density is
\begin{equation} \rho=\rho_+ \left(\frac{r_+}{r}\right)\frac{1}{\eta}
\end{equation}
where $\rho_+$ is the rest mass density at the outer event horizon radius (see
Appendix \ref{a1}).

The initial magnetic field is radial with monopole topology
\begin{equation}
\begin{array}{l} B^\phi=aB^r (A+2r)/(A\Delta+2r(r^2+a^2));\\ B^r = B_0
\sin\theta/\sqrt{\gamma};\\ B^\theta=0.
\end{array}
\end{equation} The azimuthal component is introduced in order to ensure that
the electromagnetic fluxes of energy and angular momentum vanish (see Appendix
\ref{a2}) and, thus, the initial magnetic configuration is more or less
dynamically passive.

The two-dimensional axisymmetric computational domain is 
$(r_0\!<\!r\!<\!r_1)\times(0\!<\!\theta\!<\!\pi)$,
where $r_0 = 1+0.5\sqrt{1-a^2}$ and $r_1=500$.  Notice that the inner boundary
is inside of the outer event horizon, $r_+=1+\sqrt{1-a^2}$. This can be done
because of the non-singular nature of the horizon in the Kerr-Schild
coordinates.  The computational grid is uniform in $\theta$, where it has 100
cells, and almost uniform in $\log r$, where it has 170 cells. The metric 
cell sizes are the same in both directions. The grid is divided into rings such
that for any ring the metric size of its outer cells
is twice that of its inner cells. The solution on each ring is advanced
with its own time step, the outer ring time step being twice that of the inner
ring. At the interface separating cells of different rings the conservation
of mass, energy, and momentum and magnetic flux is ensured via simple
averaging procedure.

At the inner boundary, $r=r_0$, we impose the ``non-reflective'' free-flow
boundary conditions. The use of these conditions is justified by the fact that
this boundary lies inside the outer event horizon in the region where all
physical waves move inwards. At the outer boundary, $r=r_1$ we fix the flow
parameters corresponding to the initial solution. In all runs the simulations
are terminated well before any strong perturbation generated near the centre
reaches this boundary.  At the polar axis, $\theta=0,\pi$, the boundary
conditions are dictated by the axial symmetry. Namely, quantities that 
do not vanish on the axis are reflected without change of sign, whereas 
quantities that vanish on the axis are reflected with change of sign.    

We consider models which differ only by the rotation rate of black hole, $a$,
and the mass accretion rate, $\dot{M}$.  For each value of $a$, we first deal
with the model with $\kappa\simeq 1$.  Depending on whether the BZ-mechanism is
activated in this model or not we either increase or reduce $\rho_+$ (and
hence $\dot{M}$) by the factor of three and compute another model. We proceed
in this fashion until the solution becomes qualitatively different. 
Then we compute one more model based on the mean value of $\rho_+$ for 
the previous two models. After that we select the closest two models with 
qualitatively different solutions and show them on the bifurcation diagram 
summarising the results of our study.  

\begin{figure*}
\includegraphics[width=70mm,angle=90]{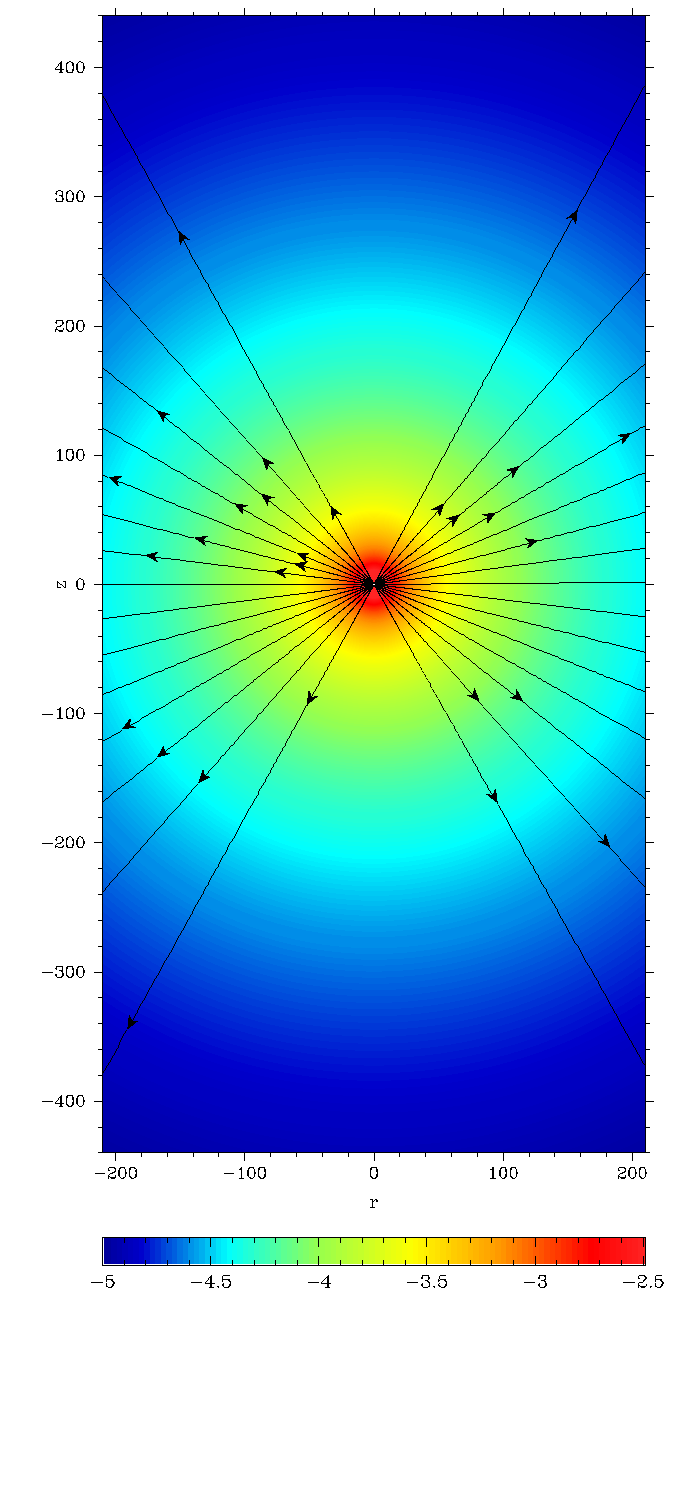}
\includegraphics[width=70mm,angle=90]{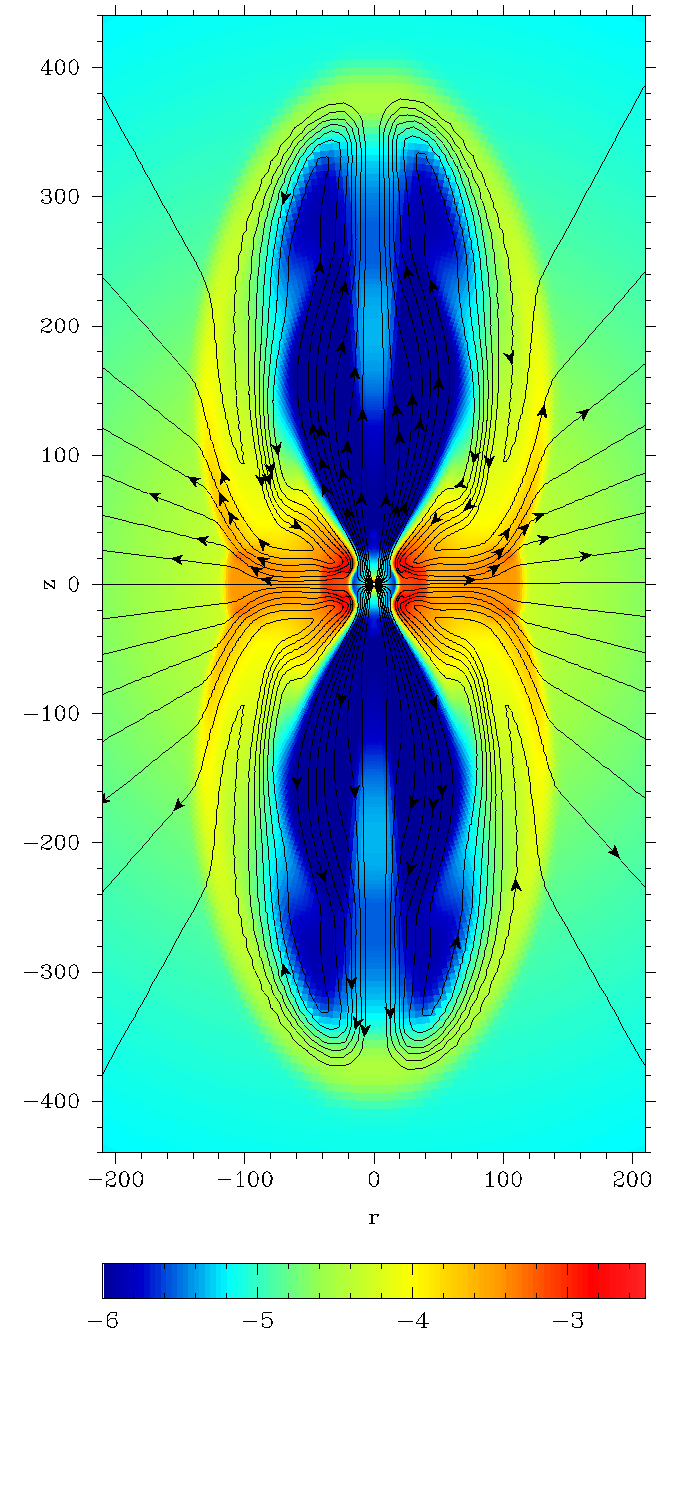}
\includegraphics[width=70mm,angle=90]{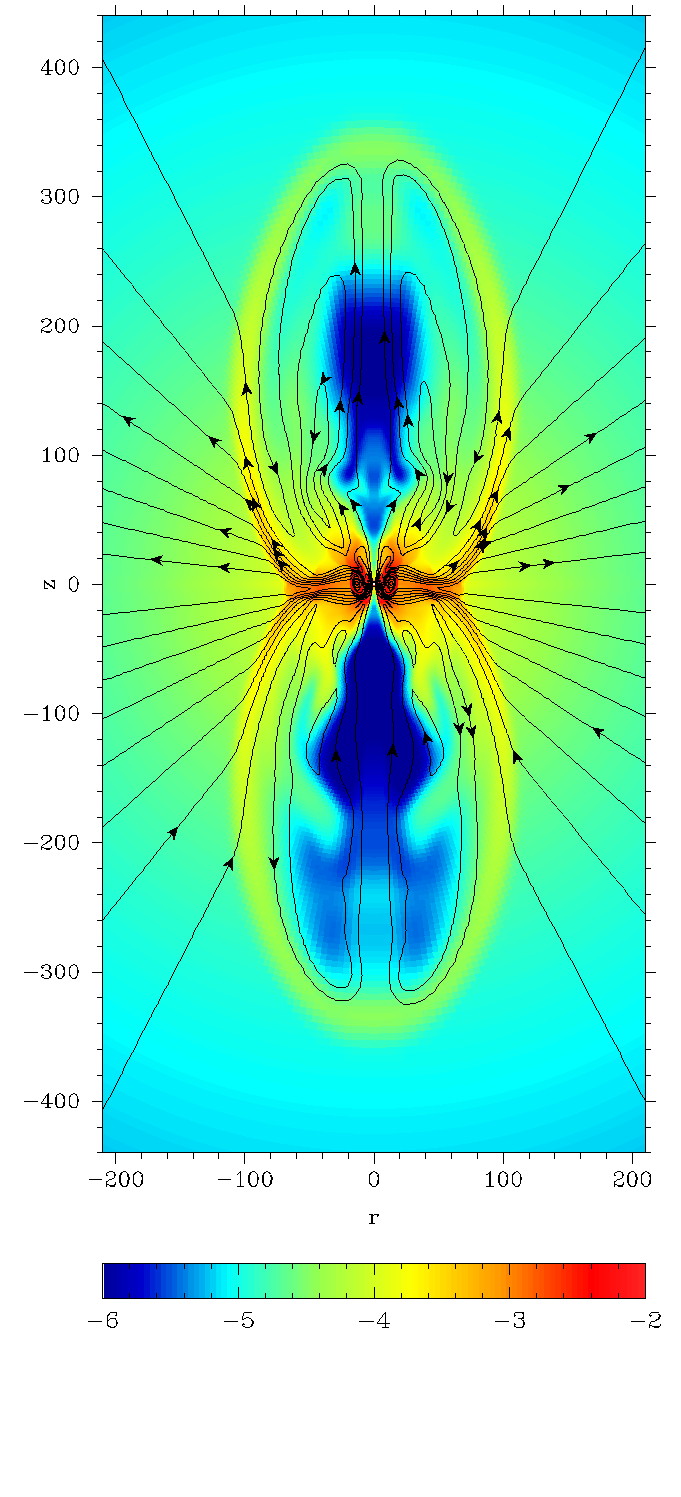}
\caption{Free-fall accretion of cold plasma with zero angular momentum onto a
Kerr black hole ($a=0.9$); the global structure.  The colour images show
$\log_{10}\rho$ in dimensionless units and the contours show the magnetic
field lines ( To be more specific the contours show the levels of poloidal
magnetic flux function, $\Psi$. At any point $(r,\theta)$ this function gives
the magnetic flux through the axisymmetric loop or radius $r$ passing through
this point.).  Top panel: model with the activation parameter $\kappa=1.2$ at
time $t=2000r_g/c$.  Middle panel: model with $\kappa=1.6$ and monopole
magnetic field at time $t=1000r_g/c$.  Bottom panel: model with $\kappa=1.6$
and split-monopole magnetic field at time $t=1600r_g/c$.  The unit of length
is $r_g$.  }
\label{sp-outer}
\end{figure*}

\begin{figure*}
\includegraphics[width=70mm,angle=90]{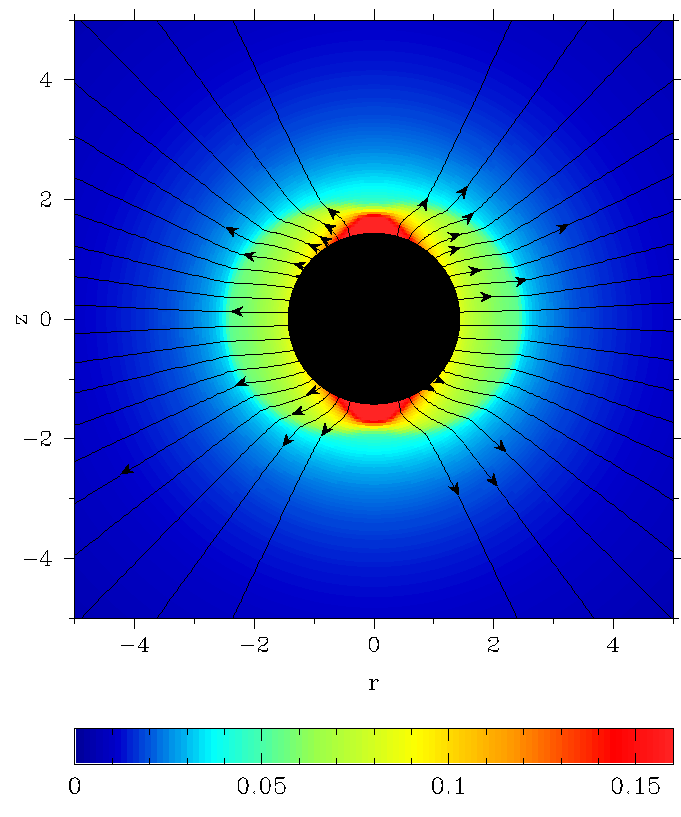}
\includegraphics[width=70mm,angle=90]{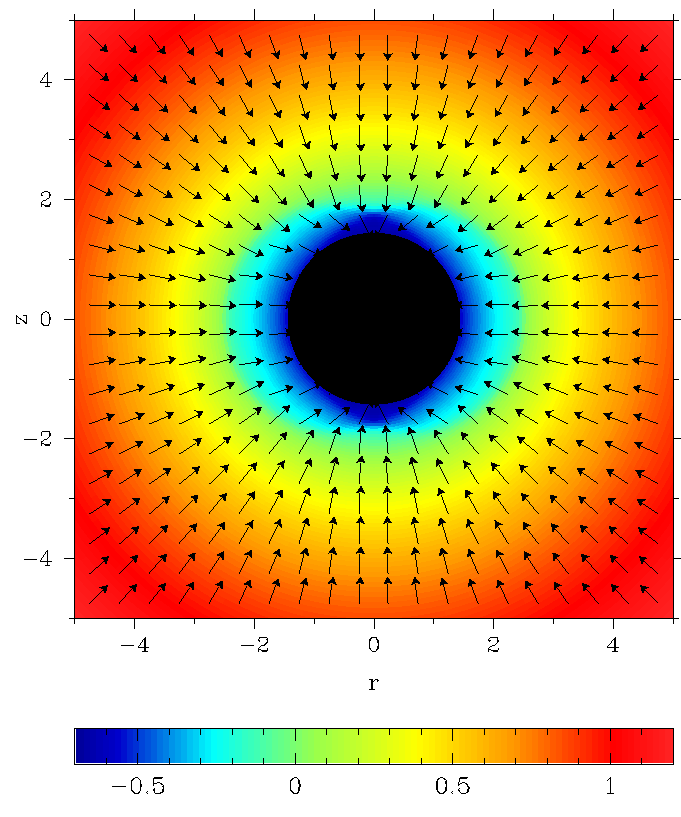}
\includegraphics[width=70mm,angle=90]{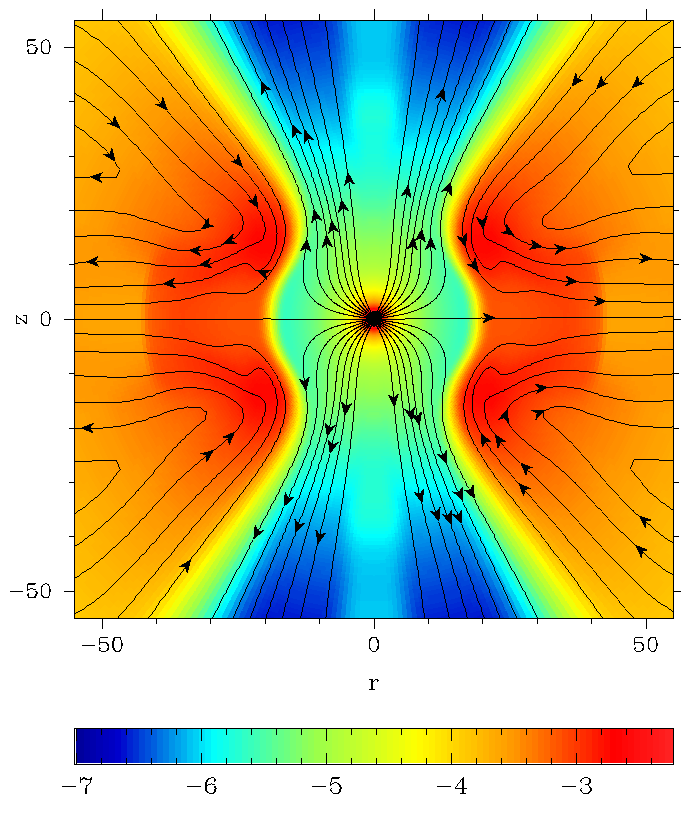}
\includegraphics[width=70mm,angle=90]{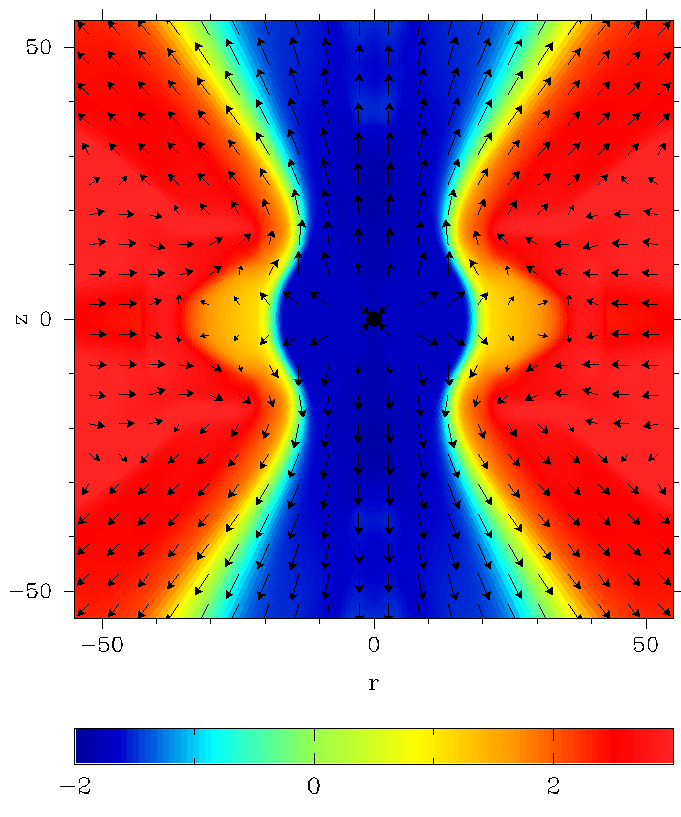}
\includegraphics[width=70mm,angle=90]{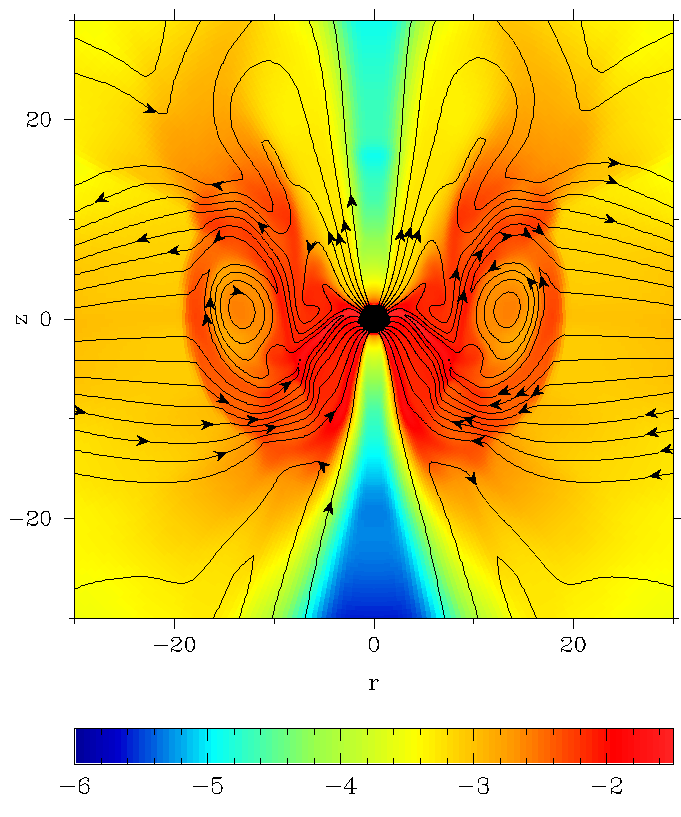}
\includegraphics[width=70mm,angle=90]{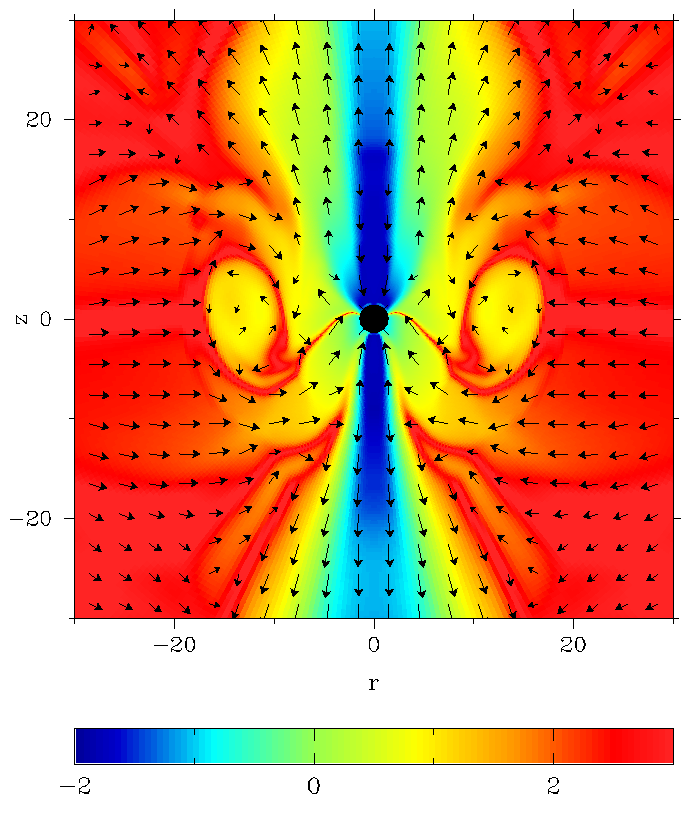}
\caption{Free-fall accretion of cold plasma with zero angular momentum onto a
Kerr black hole ($a=0.9$); the central regions.  The top panels show the model
with $\kappa=1.2$ at $t=2000r_g/c$; the middle panels show the model with
$\kappa=1.6$ and monopole magnetic field at $t=1000r_g/c$, the bottom panels
show the model with $\kappa=1.6$ and split-monopole magnetic field at
$t=1600r_g/c$.  The left panels show the rest mass density distribution
($\rho$ for the top panel and $\log_{10}\rho$ for others) and the magnetic
field lines. The right panels show the ratio of the proper rest mass and
magnetic energy densities, $\log_{10}\beta_\rho$, and the velocity field (the
length of velocity arrow is proportional to the square root of speed.).  }
\label{sp-inner}
\end{figure*}

\subsection{Results}
\label{results-1} 

The simulations do confirm the anticipated bifurcation and show that for a wide of
parameter $a$ the critical value of $\kappa$ is indeed close to unity.  
For example, the model with $a=0.9$ and $\kappa=1.2$  
is qualitatively different from the model the model with $a=0.9$ 
and $\kappa=1.6$. The first model quickly settles to a 
steady state accretion solution (see the top panels of fig.\ref{sp-outer} 
and fig.\ref{sp-inner}). In fact, on the large-scales, this solution  
is almost indistinguishable from the initial solution. 
On the small scales, near the event horizon, the differences become more 
pronounced.
In particular, a stationary accretion shock appears just outside of the 
ergosphere.  It is responsible for the jump in the rest 
mass density and the kinks of magnetic field lines seen in the top left panel 
of fig.\ref{sp-inner}. Inside the ergosphere the magnetization
parameter $\mu$ is larger than unity but only just. The magnetic field lines 
show no rotation and thus the BZ-mechanism remains completely switched-off.

In contrast, the model with $a=0.9$ and $\kappa=1.6$ exhibits  powerful
bipolar outflow which drives a blast wave into
the accreting flow (see the middle panel of fig.\ref{sp-outer}). This blast
waves overturns the accretion both in the polar and in the equatorial
direction. The black hole develops  highly rarefied magnetosphere 
(see the middle panels of fig.\ref{sp-inner}) which 
rotates with about half of 
black hole's angular velocity.
The shocked plasma of accretion flow is prevented from
accretion onto the black hole -- instead it participates in the large scale
circulations above and below the equatorial plane.  The
unphysical monopole configuration of magnetic field set for this model
prevents any escape of magnetic flux from the black hole.  This ensures
operation of the BZ-mechanism even if the confining accreting envelope is
fully dispersed into the surrounding space. 

For a more realistic dipolar configuration one would expect at least partial 
escape of magnetic flux once the outflow is developed and thus less power in 
the outflow.  However, this is unlikely to effect the
value of $\kappa_c$. Indeed, consider the case with monopole field 
where the BZ-mechanism remains switched off. Then, change the polarity of 
magnetic field lines in the northern hemi-sphere.
Because the speeds of MHD waves and the magnetic stresses remain invariant under 
change of magnetic polarity this will have no effect on the flow. 
For the very same reason, we expect 
activation of the BZ mechanism for exactly the same value of $\kappa$ in both 
magnetic configurations.
In order to verify this conclusion we computed models with the initial
magnetic field of split-monopole topology and the results fully agree with the
expectations -- the critical value of $\kappa$ remains unaffected but the
power of BZ-mechanism is reduced due partial escape of magnetic flux
from the black hole. The bottom panel of fig.\ref{sp-outer} shows
the large scale flow developed in the model with $a=0.9$, $\kappa=1.6$, and
split-monopole initial magnetic field. The flow pattern is very similar to
that of the model with monopole field but the expansion rate is noticeable
lower, approximately by the factor of 1.8 in the polar direction and even more 
in the equatorial direction. The bottom panels of fig.\ref{sp-inner} show the inner
region of this model.  One can see that the accretion is no longer fully
halted -- the accreting plasma finds its way into the black hole in the
equatorial zone and the black hole magnetosphere is confined only to the polar
region of the funnel shape.

\begin{figure}
\begin{center}
\includegraphics[width=90mm]{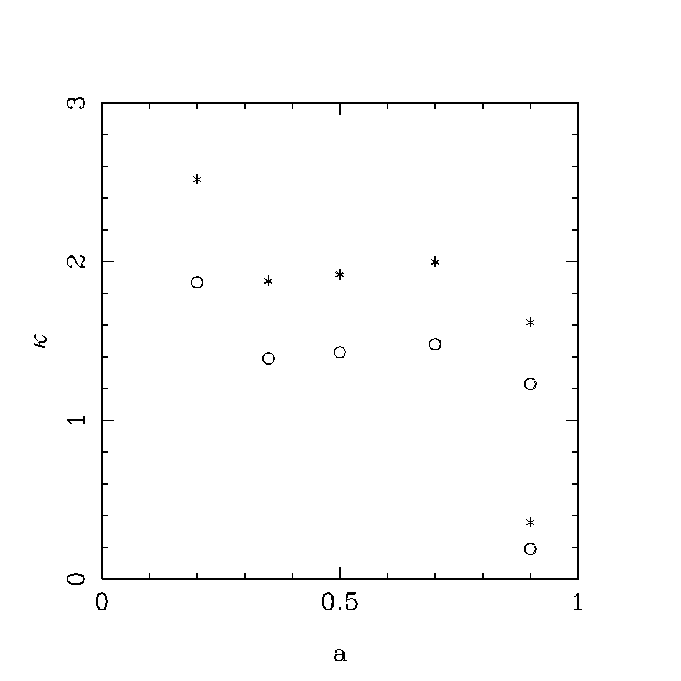}
\end{center}
\caption{The bifurcation diagram. The stars show the models with the lowest
value of $\kappa$ which still show activation of the BZ mechanism. The circles
show models with the highest value of $\kappa$ where the BZ mechanism remains
turned off.  The pair of points with $a=0.9$ in the right bottom corner of the
diagram correspond to proper collapsar simulations. The other points show the
results for test simulations.  }
\label{kappa}
\end{figure}

The results for $a=0.9$ show that not only there is a bifurcation between the
regimes with switched-on and switched-off BZ process but also that the
critical value of the control parameter $\kappa$, and hence $\beta_\rho$, is
indeed close to unity at least for rapidly rotating black holes. The results
of our parametric study, summarised in fig.\ref{kappa}, show that in fact
$\kappa_c(a)$ is a relatively weak function of $a$. It is confined within the
range $(1,2)$ for $0.2<a<1.0$.

\section{Collapsar simulations}
\label{coll-sim}

The test simulations described in Sec.\ref{test-sim} allow us to
determine the basic criteria for activation of the BZ mechanism in principle.
However, their set-up is rather artificial as in many astrophysical
applications the accreting plasma has sufficient angular momentum to ensure
formation of accretion disk and thus to render the approximation of
spherical accretion unsuitable.  This anisotropy of mass inflow is likely to
make the integral activation condition (\ref{cond-2}) less helpful.  
On the other hand, one would expect the local condition 
condition (\ref{cond-1}) to be more or less robust. In order to check this we
analysed the results obtained in our current numerical study of magnetic
collapsar model.

\subsection{Computational grid and boundary conditions} 

Like in the test
simulations we use the Kerr metric in Kerr-Schild coordinates,
$\{\phi,r,\theta\}$. We selected 
$M\sub{h}=3M_{\sun}$ and the rather optimistic value of $a=0.9$ for the 
central black hole.  
The two-dimensional axisymmetric computational domain is 
$(r_0\!<\!r\!<\!r_1)\times(0\!<\!\theta\!<\!\pi)$, with
$r_0 = (1+0.5\sqrt{1-a^2}) \; r_g= 5.4\,$km and $r_1=5700 \;r_g = 25000\,$km.
The computational grid is of the same type as in the test simulations
(Sec.\ref{setup-1}) but with more cells, $450\times180$.  The boundary
conditions are also the same as in the test simulations.

\subsection{Microphysics} 
For these simulations we use realistic equation of state (EOS)  
that takes into account the contributions from radiation, lepton gas including
pair plasma, and non-degenerate nuclei (hydrogen, helium, and oxygen). 
This is achieved via incorporation of the EOS code HELM \citep{TS00},  
which can be downloaded from the web-site 
http://www.cococubed.com/code\_pages/eos.shtml.  

The neutrino cooling is computed assuming optically thin regime and takes 
into account URCA-processes \citep{IIN69}, pair annihilation, photo-production, 
and plasma emission \citep{S87}, as well as synchrotron neutrino 
emission \citep{B97}. In fact, URCA-processes strongly dominate over other 
mechanisms in this problem.  
Photo-disintegration of nuclei is included via modification of EOS following 
the prescription given in \citet{ABM05}.  The equation for mass
fraction of free nucleons is adopted from \citet{WB92}. We have not included 
the radiative heating due to annihilation of neutrinos and antineutrinos
produced in the accretion disk mainly because this requires elaborate and time 
consuming calculations of neutrino transport.

\subsection{Model of collapsing star}

The collapsing star is introduced using the simple free-fall model by
\citet{Bethe}. In this model it is assumed that immediately before the collapse the
mass distribution in the star satisfies the law $\rho r^3=$const and that the
infall proceeds with the free-fall speed in the gravitational field of the central 
black hole.  This gives us the following radial velocity
\begin{equation} v^r\sub{ff}=(2GM/r)^{1/2}
\label{b1}
\end{equation} 
and mass density
\begin{equation} \rho = C_1\times10^7 \left(\frac{t_s}{1\mbox{~s}}\right)^{-1}
\left(\frac{r}{10^7 \mbox{cm}}\right)^{-3/2} \mbox{g~cm}^{-3}
\label{b2}
\end{equation} 
for the collapsing star, where $C_1$ is a constant that depends on the star 
mass \citep{Bethe} and $t_s$ is the time since the start of collapse. 
The corresponding accretion rate and ram pressure are
\begin{equation} \dot{M} = 0.056~C_1 \left(\frac{M}{3M_{\sun}}\right)^{1/2}
\left(\frac{t_s}{1\,\mbox{s}}\right)^{-1} M_{\sun} \mbox{s}^{-1},
\label{b3}
\end{equation}
\begin{equation} p_{\mbox{\tiny ram}} = 7.9\times 10^{26} C_1
\frac{M}{3M_{\sun}} \left(\frac{t_s}{1\mbox{~s}}\right)^{-1}
\left(\frac{r}{10^7\mbox{cm}}\right)^{-5/2} \frac{\mbox{g}}{\mbox{cm~s}^{2}}
\label{b4}
\end{equation} 
respectively.  For a core of radius $r_c=10^9$cm and mass
3.0$M_{\sun}$ the collapse duration can be estimated as
$t_c=2r_c/3v_{\mbox{ff}}\simeq 0.6$s. Since, in this study we explore the
possibility of early explosion, that is soon after the core collapse, we set
$t_s=1$s. Because GRBs are currently associated with more massive progenitors
we consider $C_1=3,9$.  This gives us the accretion rates of $ \dot{M}
\simeq 0.166 M_{\sun} \mbox{s}^{-1}$ and $ \dot{M} \simeq 0.5 M_{\sun}
\mbox{s}^{-1}$ respectively.

On top of this we endow the free-falling plasma with angular momentum and
poloidal magnetic field.  The angular momentum distribution is
\begin{equation} l = \sin\theta \left\{
\begin{array}{lcl} l_0(r\sin\theta/r_l)^2 &\text{if}& r\sin\theta<r_l \\ l_0
&\text{if}& r\sin\theta>r_l
\end{array} \right. ,
\label{b5}
\end{equation} 
where $r_l = 6300$km and
$l_0=10^{17}\mbox{cm}^2\mbox{s}^{-1}$.\footnote{ The intention was to set up a
solid body rotation within the cylindrical radius $r_l$, however, by mistake
an additional factor, $\sin\theta$, was introduced in the start-up models (The
same mistake was made in the simulations described in \citet{BK08}.)
Although somewhat embarrassing, this does not affect the main results of this
study.}

\subsection{Initial magnetic field} 

The origin of magnetic field is probably
the most important and difficult issue of the magnetic model of GRBs. One
possibility is that this field already exists in the progenitor and during the
collapse it is simply advected onto the black hole. The recent numerical
studies indicate that the poloidal and the azimuthal components of the
progenitor field should have similar strengths \citep{BS04}. The magnetic flux
freezing argument then suggests that during the collapse the poloidal field
becomes dominant and thus in our simulations we may ignore the initial
azimuthal field.  As to the initial poloidal component we adopt the  
solution for a uniformly magnetized sphere in vacuum\footnote{ 
Thus, we ignore the magnetic field that is already threading the black hole by 
the start of simulations.}. To be more specific, we assume 
that inside the sphere of radius $r_m=4500\,$km  the magnetic field is uniform 
and has strength of either $B_0=3\times10^{9},\,10^{10},$ or $3\times 10^{10}$G, 
whereas outside it is described by the solution for magnetic dipole. 
The selected values for the field strength allow to capture the BZ bifurcation.  
Given the limitations of axisymmetry we can only consider the case of
aligned magnetic and rotational axes. 

\begin{figure*}
\includegraphics[width=60mm,angle=0]{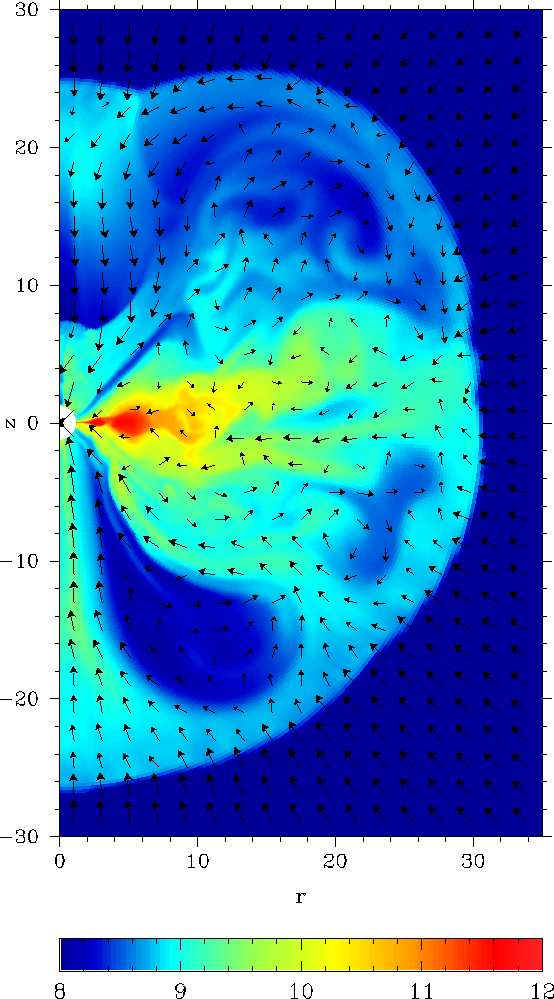}
\includegraphics[width=60mm,angle=0]{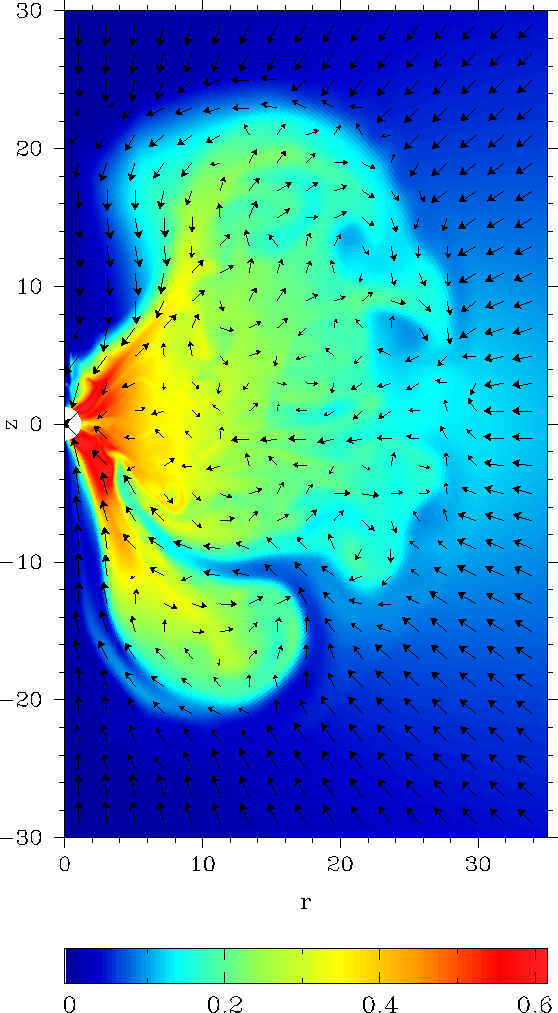}
\includegraphics[width=60mm,angle=0]{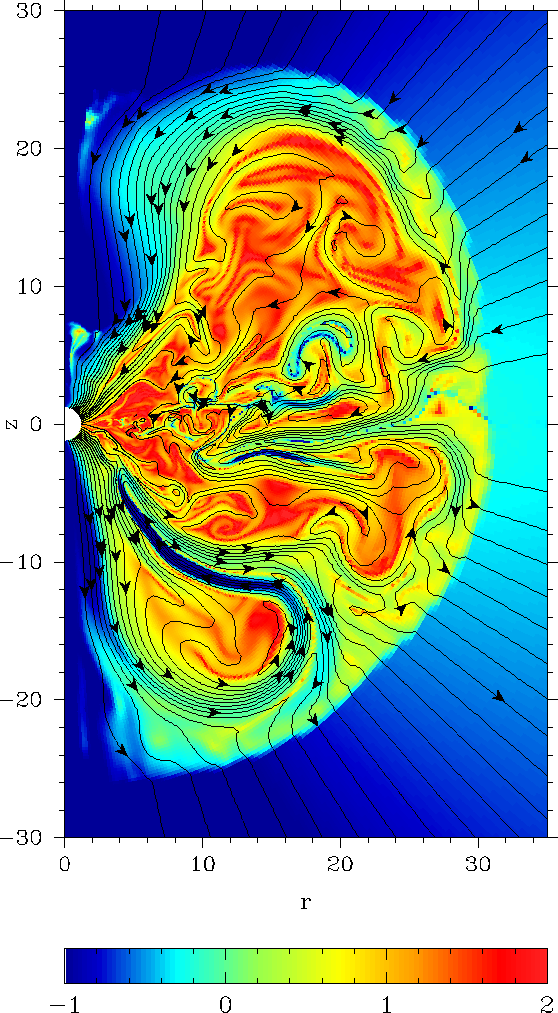}
\includegraphics[width=60mm,angle=0]{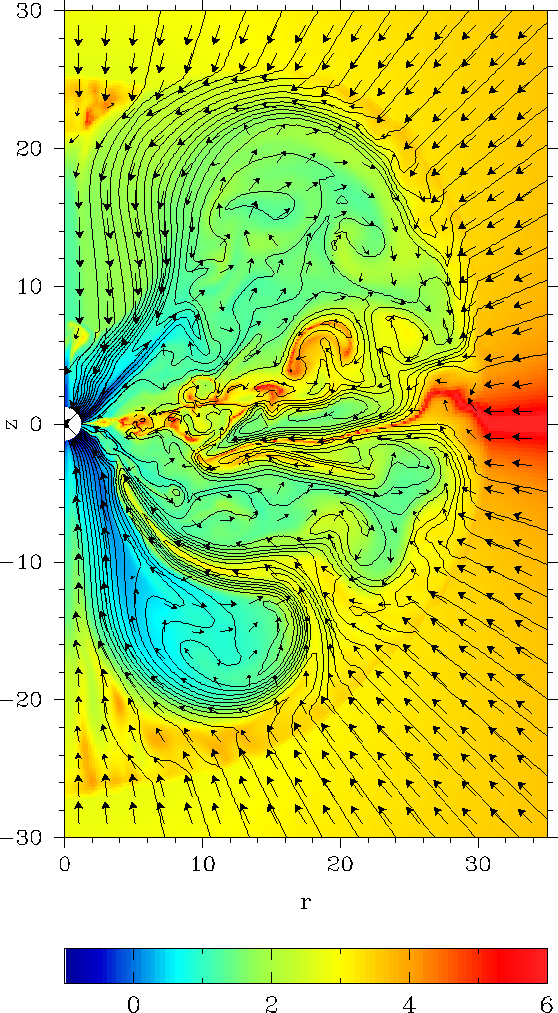}
\caption{Model B close to the explosion, at $t=237$ms.  Top left:
$\log_{10}\rho$ and the poloidal velocity; Top right: the angular velocity of
plasma and the poloidal velocity; Bottom left: the ratio of azimuthal and
poloidal components of magnetic field, $\log_{10}(|\bB_\phi|/|\bB_p|)$, and
the magnetic field lines; Bottom right: $\log_{10} \beta_\rho$, the magnetic
field lines, and the poloidal velocity field.  The length of velocity arrows
is proportional to the square root of speed.  }
\label{before}
\end{figure*}

\subsection{Results}
\label{results-2}

\begin{figure}
\includegraphics[width=73mm,angle=0]{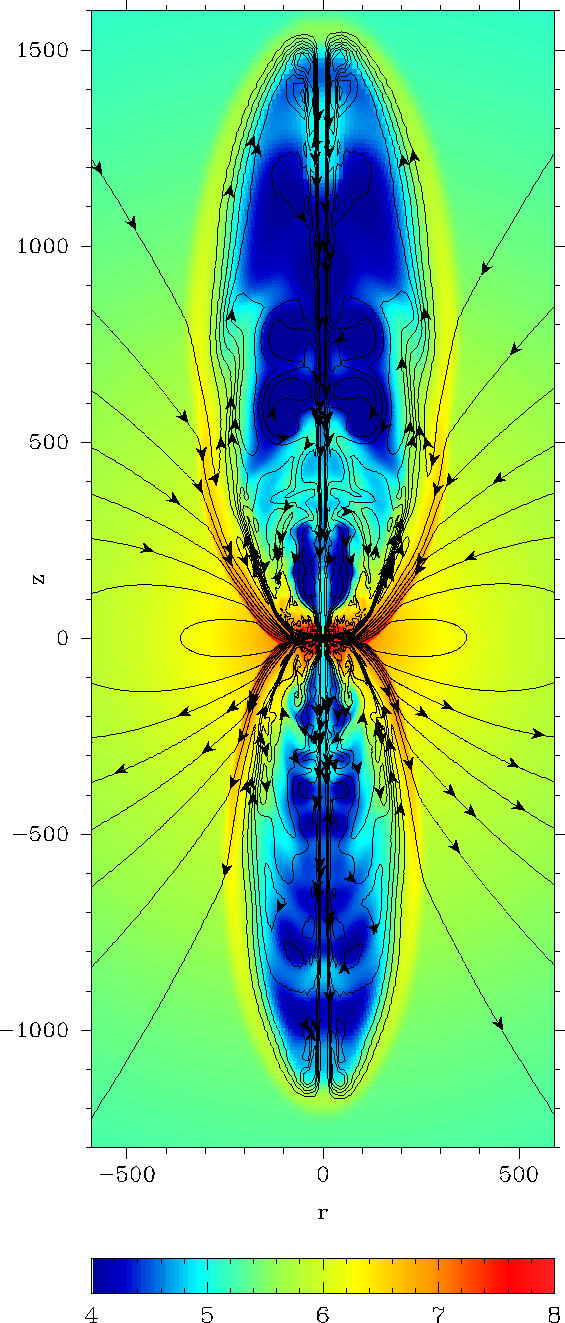}
\caption{Model B soon after the explosion.  The figure shows the solution at
$t=332$ms. The colour image gives $\log_{10}\rho$ and the contours show the
magnetic field lines.  }
\label{after}
\end{figure}

Table \ref{model-tab} describes the key parameters of six models investigated
in this study and also summarises the outcome of simulations. One of the
parameters, $\Psi^{max}$, is the total magnetic flux enclosed by the
equator of the uniformly magnetized sphere in the initial solution.  
Obviously, this gives us the highest magnetic flux that can be
accumulated by the black hole in the simulations. Whether this maximum value
is actually reached at some point depends on the progenitor rotation. For slow
rotation the whole of the uniform magnetized core can be swallowed by the hole
before the development of accretion disk. At this point $\Psi_h$ equals to
$\Psi^{max}$ and then it begins to decline as the result of advection of the
oppositely directed parts of dipolar loops (These loops are clearly seen in
fig.\ref{after}).  For fast rotation the accretion disk forms before $\Psi_h$
reaches $\Psi^{max}$ and after this point the growth of $\Psi_h$ slows down
noticeably as the disk accretion is much slower compared to the free-fall
accretion. Once the oppositely directed parts of magnetic loops begin to cross
the accretion shock and enter the turbulent domain that surrounds the disk the
annihilation of poloidal magnetic field in this domain becomes an important
factor. As the result the maximum value of magnetic flux available in the
solution begins to decrease before $\Psi_h$ reaches $\Psi^{max}$.

\begin{table}
\begin{tabular}{|c c c c c c c c|} \hline name & $B\sub{0,10}$ & $C_1$ &
$\Psi\sub{27}^{max}$ & $\kappa_{max}$ & Exp & $\Psi\sub{h,27}$ &
$\dot{E}\sub{51}$ \\ \hline A & 3 & 3 & 19.1 & 1.07 & Yes & 11.6 & 9.6\\ B & 3
& 9 & 19.1 & 0.62 & Yes & 13.1 & 12.7\\ C & 1 & 3 & 6.67 & 0.36 & Yes & 2.80 &
0.4\\ D & 1 & 9 & 6.67 & 0.21 & No &&\\ E & 0.3 & 3 & 1.91 & 0.11 & No &&\\ F
& 0.3 & 9 & 1.91 & 0.06 & No &&\\ \hline
\end{tabular}
\caption{Summary of numerical models in collapsar simulations.  $B_{0,10}$ is
the initial magnetic field strength in units of $10^{10}$G, $C_1$ is the
density parameter in eq.\ref{b2}, $\Psi\sub{27}^{max}$ is the maximum value of
magnetic flux that can be accumulated by the black hole in units of
$10^{27}\mbox{G}\,\mbox{cm}^2$, $\kappa_{max}$ is the value of $\kappa$
corresponding to $\Psi\sub{27}^{max}$, ``Exp'' is the explosion indicator,
$\Psi\sub{h,27}$ is the actual magnetic flux accumulated by the black hole by
the time of explosion, $\dot{E}\sub{51}$ is the mean power of the BZ mechanism
in units of $10^{51}$erg/s.  }
\label{model-tab}
\end{table}

$\kappa_{max}$ is the value of parameter $\kappa$ computed using $\Psi^{max}$
instead of the actual value of black hole flux in eq.\ref{cond-2}. Quick
inspection shows that the critical value of $\kappa$ based on the values of
$\kappa_{max}$ is about $\kappa_c=0.3$ (see models C and D in
table~\ref{model-tab}.), almost five time lower than that in the test
simulations of spherical collapse. In fact, the actual value of magnetic flux 
accumulated by the black hole, $\Pi_h$, by the time of explosion is always noticeably 
lower than $\kappa_{max}$ (see table~\ref{model-tab}).  In model C it is 2.4 times 
lower thus driving the actual critical value of $\kappa$ down to the value around 0.1.
This result is undoubtedly connected to the highly anisotropic
nature of the flow in the close vicinity of the black hole that develops in
the simulations. Most importantly, there is a clear separation of the mass and
magnetic fluxes: while a large fraction of mass is accreted through the
accretion disk, the bulk of magnetic field lines connected to the black hole
occupy the space above (and below) the disc (see fig.\ref{before}).

All models considered in this study first pass through the passive phase of
almost spherical accretion. During this phase the black hole accumulates
magnetic flux but as $\Psi^{max}$ is not sufficiently high the
Blandford-Znajek mechanism remains switched-off.  Then, as the angular
momentum of plasma approaching the black hole exceeds that of the marginally bound 
orbit, the accretion disk forms in the equatorial plane and an accretion shock
separates from its surface. This shock is best seen in the density plot of
fig.\ref{before} where it appears as a sharp circular boundary between the
light blue and dark blue areas of radius $\simeq 30r_g \simeq 135$km.  In
fact, the accretion shock oscillates in a manner reminding the oscillations
found earlier in supernova simulations \citep[e.g.][]{BMD03,BJRK06,SJFK08}, as 
well as in simulations of accretion flows onto black holes \citep{NY09}, 
and attributed to the so-called Stationary Accretion Shock Instability (SASI). 
Most likely, the physical origin of shock oscillations in our simulations is the 
same but we have not explored this issue in details.

During this second phase the differential rotation results in winding up of
the magnetic field in the disk and its corona where the azimuthal component
of magnetic field begins to dominate (top-right and bottom-left panels of
fig.\ref{before}).  The corona acts as a ``magnetic shield'' deflecting the
plasma coming at intermediate polar angles towards either the equatorial plane
or the polar axis (see fig.\ref{before}).  In models D,E and F the second phase
continues till the end of run and the Blandford-Znajek process remains
switch-off permanently.  There are no clear indications suggesting that this
phase may end soon although the accretion disk grows monotonically in radius
and this may eventually bring about a bifurcation.

In models A,B,C after several oscillations the solution enters the third phase
which is characterized by a non-stop expansion of the accretion shock which
then turns into a blast wave. During this transition the black hole develops a
low mass density magnetosphere which occupies the axial funnel through which
the rotational energy of the black hole is transported away away in the form 
of Poynting flux and powers the blast wave.  The typical large-scale structure of 
the the solution during the
third phase is shown in fig.\ref{after} (see also \citet{BK08} ).  It is
rather similar to the structure observed in the simulations of spherical
accretion in the case of split-monopole magnetic field (see
Sec.\ref{results-1}).
        
Figure \ref{before} shows the solution for model B at time $t=237$ms, which is
close to the end of the second phase, and helps to understand how exactly the
BZ mechanism is activated in our simulations.  At this stage, the accretion
disk is well developed and accounts for $\simeq 50-60\%$ of mass flux through
the event horizon whereas most of the magnetic field lines threading the
horizon avoid both the disk and corona that surrounds the disk and contains
predominantly azimuthal magnetic field.  The strong turbulent motion in the
disk and its corona must be the reason for this expulsion of magnetic flux
\citep{Z57,R68,T98}.  In any case, the observed separation of magnetic and mass 
fluxes results in a rather inhomogeneous magnetization of plasma near the black hole. 
Most importantly, there are regions that have much higher local magnetization compared 
to the case of spherical accretion with the same value of integral control 
parameter $\kappa$. 

Moreover, the magnetic shield action of the disk corona can promote further
local increase of magnetization. Indeed, the corona can deflect plasma
passing through the accretion shock at intermediate latitudes in the direction
which does not
necessarily coincide with the local direction of the magnetic field.  For
example, in figure \ref{before} the stagnation point of post-shock flow in
the upper hemisphere is located at $(r,z)\simeq (25,20)r_g$ whereas the
separation between the magnetic field lines connected to the hole and to the
disk occurs at much lower latitude, around $(r,z)=(28,10)r_g$.  Thus, the
direct mass loading of the field lines entering the shock between these points
with is currently suspended.  On the other hand, at the other end of these field lines, where
they penetrate the event horizon, their unloading continues as usual. As
the result, the magnetization along these lines increases allowing to 
reach the local criticality condition, $\beta_\rho <1 $ (see the intense blue 
regions in the left panel of fig.\ref{before}). Once this condition is satisfied
the BZ mechanism turns on, at this point only very locally and the initial rotation
rate of the field lines is significantly lower then $0.5\Omega_h$. However,
an additional magnetic energy is now being pumped into the shield via the BZ
mechanism promoting its expansion and increasing its efficiency as a flow deflector.  
This causes further increase of the magnetization at its base, further increase
of magnetosphere's rotation rate and thus higher BZ power and so on. The process
develops in a runaway fashion.

\begin{figure}
\begin{center}
\includegraphics[width=90mm]{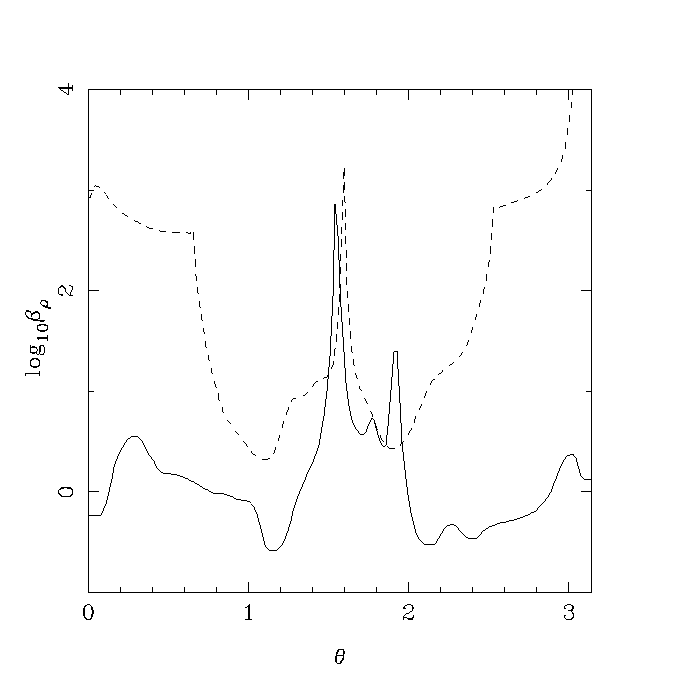}
\end{center}
\caption{The magnetization parameter $\beta_\rho=4\pi\rho c^2/B^2$ at $r=2r_g$
for models B at $t=237$ms (solid line) and D at $t=360$ms (dashed line).  }
\label{loc-cond}
\end{figure}

Figure \ref{loc-cond} shows the magnetization parameter $\beta_\rho$ at
$r=2r_g$ as a function of the polar angle for model B (at the same time as in
fig.\ref{before}) and for model D at a similar time.  One can see that for
model D the magnetization is lower. In fact, the local criticality condition,
$\beta_\rho<1$, is not satisfied anywhere for this model and this explains why
the explosion is not produced in this model\footnote{ These results also
suggest that both numerical diffusion and resistivity reduce chances of
successful explosion in computer simulations via smoothing the distributions of
mass and magnetic field and thus decreasing inhomogeneity of $\beta_\rho$.}.

\section{Discussion}
\label{disscusion}

\subsection{The strength and origin of magnetic field.}
\label{magnetic}

There seems to be only two possible origins for the magnetic field in the
collapsar problem. Either this is a fossil field of the progenitor star of the
field generated in the accretion disk and its corona. Here we consider both
possibilities.

The blunt application of the integral condition (\ref{cond-2}) for activation
of the Blandford-Znajek mechanism gives us the critical value of magnetic flux
accumulated by the black hole
$$
\Psi_{h,c}=4.3\times10^{28} \left( \frac{M_h}{3M_\odot} \right) \left(
\frac{\dot{M}\mbox{s}}{M_\odot} \right)^{1/2} \!\!\mbox{G}\,\mbox{cm}^2.
$$ 
Although the strength of magnetic field in the interior of massive stars which
end up as failed supernovae is not known this is uncomfortably high compared to 
fluxes deduced from the observations of ``magnetic stars'' and rises
doubts about the fossil hypothesis. However, the actual values of the critical
magnetic flux found in our simulations are significantly lower. As we have
already commented in Sec.\ref{coll-sim} this is due to the anisotropy 
of accretion in the collapsar model. This anisotropy is expected to increase 
with time as more distant parts of the progenitor with higher angular momentum 
enter the accretion shock (we could not study this late phase of the collapse,
$t\gg1\,$s, because of various technical issues.).  
Due to the higher angular momentum the predominant motion downstream
of the shock will be towards the equator and this should result in (i)
further reduction of the mass flux and the ram pressure of accreting plasma in
the polar region and (ii) more effective mass-unloading of magnetic field lines
entering the black hole (see Sec.\ref{results-2}). Both effects promote 
higher plasma magnetization in the polar region around the black hole which may
eventually exceed the local criticality condition (\ref{cond-1}) even in
models with $\Psi_h < 10^{27}\mbox{G}\mbox{cm}^2 $, making
the fossil hypothesis more plausible.

Our results seem at odds with the Newtonian simulations by \citet{PMAB03} who 
obtain magnetically driven explosions powered by the disk wind already within 
$300\,$ms in physical time since the start of simulations for a model with 
similar mass accretion rate, $\dot{M}\simeq 0.29M_\odot/$s, but much lower 
magnetic field, $\Psi\simeq 10^{26}\mbox{G}\,\mbox{cm}^2$. Since their initial model 
is similar to that of \citet{MW99} the difference cannot be attributed to the development 
of highly rarefied polar channel seen in the simulations of \citet{MW99} as this occurs 
only several seconds later. 
Other possible reasons include differences in the framework, 
details of the initial solutions, resolution, implementation of 
microphysics\footnote{
In particular, \citet{PMAB03} state that neutrino cooling is dominated by 
pair annihilation whereas we find that it is the lepton capture on free 
nucleons which is dominant in our models. } etc. 
Hopefully, simulations by other groups will soon clarify this issue.

Another factor that is likely to reduce the constraint on the strength of the
fossil field, and the one that is completely ignored in our simulations, is the
plasma heating due to neutrino-antineutrino annihilation.  In the most extreme
scenario this heating completely overpowers accretion in the polar direction
and drives a relativistic GRB jet \citep{MW99}. Although this is an
attractive model of GRB explosions by itself we also note that the creation of
low density funnel in this model helps to activate the BZ mechanism along the
magnetic field lines that are advected inside this funnel. Thus, the neutrino heating
and BZ mechanism may work hand in hand in driving GRBs and hypernovae.

Even in failed supernovae the stellar core collapse does not lead directly to 
formation of a black hole but still proceeds through the 
proto-neutron star (PNS) phase first. If the period of PNS is very short, around several 
millisecons, its rotational energy is sufficient to strongly modify the direction 
of collapse and the initial conditions for the black hole phase.  
In particular, the magnetic stresses due to either fossil or MRI-generated magnetic 
field can drive powerful bipolar outflows \citep{MBA06,OAM06,BD07}. As reported by 
\citet{BD07} these outflows can coexist with equatorial accretion and may not be able 
to prevent eventual collapse of PNS. However, they certainly overcome accretion 
in the polar direction and create favorable conditions for activation of the 
BZ mechanism. The very fast rotation of progenitor's Fe core required in this 
scenario, with the period around few seconds, is unlikely to be achieved in stars 
with strong fossil magnetic field (see Sec.\ref{rotation}).

A related issue is whether a significant fraction of fossil magnetic flux 
available in the progenitor can actually be accumulated by the black hole. 
The traditional viewpoint is that the inward advection of
magnetic field in accretion disk is effective and can result in build-up of
poloidal magnetic field up to the equipartition with the disk gas pressure
\citep{BR74,BR76,BZ77,MT82,BBR84}.  On the other hand, it has also been argued
that in standard viscosity driven accretion disk the turbulent outward
diffusion of poloidal field may easily neutralise the inward advection well
before it reaches the equipartition strength\citep{LPP94,HPB96,GA97,LOP99}.
However, the disk angular momentum can also be removed by magnetic stresses 
via interaction with the disk corona and wind and
this may change the balance in favour of the inward accretion of
poloidal magnetic flux \citep{SU05,RL08}. This has been demonstrated in recent
3D Newtonian simulations of radiatively inefficient accretion disks where the
poloidal magnetic field reached the equipartition strength near the central
``black hole''\citep{I08}.  In our simulations, a significant fraction of
black hole's magnetic flux is accumulated before the development of accretion
disk. However, the flux does not escape after the disk is formed. On the
contrary, it keeps increasing until the oppositely directed magnetic field
lines begin to enter the accretion shock.  Thus, our results are in agreement with those by
\citet{I08} and suggest strong magnetic interaction between the disk and its
corona/wind.
   
The alternative origin of magnetic field is via dynamo process in the
turbulent accretion disk. In this case the saturated equilibrium magnetic
field strength can be estimated from the energy equipartition argument
\begin{equation} \frac{B_t^2}{8\pi} \simeq \rho\frac{v_t^2}{2},
\end{equation} where $B_t$ and $v_t$ are the turbulent magnetic field and
velocity respectively.  However, the typical scale of this field is rather
small, of the order of the disk hight, $h \ll r_g$.  
When magnetic loops of such small scale are
advected onto the black hole one would expect quick, on the time scale of
$r_g/c$, reconnection and annihilation of magnetic field, resulting
in low residual magnetic flux penetrating the black hole. \citet{TP96} argued
that similar reconnection processes in the disk corona may lead to inverse
cascade of magnetic energy on scales exceeding $h$.  The expected spectrum in
this range is $B_\lambda \propto 1/\lambda$ and thus on the scale of $r_g$ one could find
the regular magnetic field of strength
$$
  B_p \simeq \frac{h}{r_g}B_t.
$$   
The standard theory of $\alpha$-disks gives
$$
  v_t\simeq\alpha\Omega_k h,
$$
where $\Omega$ is the Keplerian frequency, $h$ is the half-thickness of the
disk and $\alpha=0.1\div0.01$ is a free parameter of this theory.  According
to the theory of thin neutrino cooled accretion disks by \citet{PWF99}
$$
\frac{h}{r} = 0.09 \!\left(\frac{\alpha}{0.1}\right)^{\!0.1}
\!\!\left(\frac{M_h}{M_\odot}\right)^{\!-0.1}
\!\!\left(\frac{r}{r_g}\right)^{\!0.35}
$$
and
$$
\rho = 2.4\times10^{15} \!\left(\frac{\alpha}{0.1}\right)^{\!-1.3}
\!\!\left(\frac{M_h}{M_\odot}\right)^{\!\!-1.7}
\!\!\!\left(\frac{r}{r_g}\right)^{\!\!-2.55}
\!\!\!\left(\frac{\dot{M\mbox{s}}}{M_\odot}\right)
\frac{\mbox{g}}{\mbox{cm}^3}.
$$
Putting all these equations together we find the typical black hole magnetic
flux
\begin{equation} \Psi_h= 3\times10^{26}
\!\left(\frac{\alpha}{0.1}\right)^{\!0.55}
\!\!\left(\frac{M_h}{M_\odot}\right)^{\!0.95}
\!\!\left(\frac{\dot{M\mbox{s}}}{M_\odot}\right)^{\!0.5}
\mbox{G}\,\mbox{cm}^2.
\label{mflux1}
\end{equation} 
The corresponding power of BZ mechanism is 
\begin{equation} 
\dot{E}\sub{BZ} \simeq 1.3\times10^{50} f_2(a) 
\left(\frac{\alpha}{0.1}\right)^{1.1} 
\left(\frac{\dot{M}\mbox{s}}{M_\odot}\right) 
\,\mbox{erg}\, \mbox{s}^{-1}
\label{e-bz1}
\end{equation}
almost independently on the black hole mass. One can see that
even for maximally rotating black holes one should not expect 
much more than $10^{51}$erg from a collapse of a $10M_\odot$ 
progenitor. On the other hand, in this model one should expect 
a much more powerful magnetically driven wind from the accretion disk 
\citep{LOP99}. 
Mass loading of this wind is likely to be much higher due to the processes 
similar to solar coronal mass injections are the terminal Lorentz factors 
are not expected to be as high as required from the GRB observations. 
However, the disk wind could be responsible for the $\sim\! 10^{52}$ergs
required to drive hypernova.     
Having said that, we need to warn that the above equations are purely Newtonian 
and thus their application to the very vicinity of black hole may introduce large
errors. Moreover, the current theory of magnetic dynamo in accretion disks can
hardly be called well developed and so one could expect few surprises 
as our understanding of the dynamo improves. Finally, the above picture is 
in conflict with our numerical models, particularly those which 
do not show BZ driven explosions. It is difficult to pin down the exact 
reasons for this but the key issues are likely to be the condition of 
axisymmetry and numerical resolution.

\subsection{The nature of rotation} 
\label{rotation}

With a typical rotation rate of $200\,$km/s on the zero-age main-sequence
\citep[e.g.][]{P96,H97} the standard evolutionary models of solitary stars
predict very rapid rotation of stellar cores by the time of collapse
\citep{HLW00,HMM04,HMM05}.  According to these models the typical periods of
newly born pulsars in successful supernovae are around one millisecond 
and ``failed supernovae'' conceive rapidly rotating black holes with 
accretion disks, as required for GRBs. These results, however, 
contradict to observations which clearly
indicate that young millisecond pulsars and GRBs are much more rare.

When the magnetic torque due to the relatively weak magnetic field generated
via the Tayler-Spruit dynamo \citep{T73,S02} is included in the evolutionary models
they result in significant spin down of stellar cores during the red giant
phase and during the intensive mass loss period characteristic for massive
stars at the Wolf-Rayet phase \citep{HWS05}.  As the result the predicted
rotation rates for newly born pulsars come to agreement with the observations
but on the other hand, the core rotation becomes too low compared to what is
required for the collapsar model of GRBs \citep{HWS05}.  Recently, it was
proposed that low metalicity could be the factor that allows to reduce 
the efficiency of magnetic braking in GRB progenitors \citep{WH06,YCL06}.  
On one hand, the mass-loss rate of WR-stars decreases significantly with metalicity. 
On the other hand, the evolutionary models of \citet{WH06} show that low metalicity 
stars rotating close to
break-up speed may become fully convective, chemically homogeneous, and by-pass
the red giant phase.  This result has been confirmed by \citet{YCL08} but
their models also show that even stars with metalicity two thousand times
below that of Sun inflate above 50 solar radii by the stage of carbon
burning. For a potential progenitor of even long duration GRBs the
corresponding light-crossing time is uncomfortably high, $t\sub{ls}>120$s. 
Indeed, this 
makes very difficult to explain the large observed fraction of long GRBs
with the gamma ray burst duration of only 3-30 seconds. It is somewhat curious
but the high mass loss rates of massive stars with solar metalicity has long
been considered as an important factor in the evolution of GRB progenitors as
it can result in the formation of a compact helium star (WR-star) via total
loss of extended hydrogen envelop \citep{W93}.  Obviously, these complications
are not specific to the standard fireball model of GRBs. In fact, the magnetic
field of magnetic stars is much stronger than that predicted by the
Taylor-Spruit theory of stellar dynamo and thus the magnetic braking of GRB
progenitors is more severe in our magnetic model of GRB central engine which
relies on strong fossil magnetic field.\footnote{ The magnetic field of
magnetic stars is most likely to be inherited from the interstellar medium
during star formation~\citep{M87,BS04}.}

To a large extent the above complications are specific to solitary stars and
can be avoided in close binaries. Firstly, the extended envelope of a giant
star is only weekly bounded by gravity and is easily dispersed in close
binary, predominantly in the orbital plane \citep{TaS00}.  Secondly, even in
the case of high metalicity and/or strong magnetic field the stellar rotation
rate may remain high due to the synchronisation of spin and orbital
motions. Indeed, the minimum separation between the components of binary
star, $L_c$, is determined by the size of the critical Roche surfaces as
\begin{equation} 
L_c \simeq 2.64 R q^{0.2},
\end{equation} 
where $R$ is the radius of more massive component and $q$ is
the ratio of component masses $(q\le1)$. The rotational period is given by
\begin{equation} 
\Omega^2=(1+q)\frac{GM}{L^3},
\end{equation} 
where $L$ is the actual separation of component and $M$ is the
mass of more massive component. Using $q=1$, $L=3R$, $R=5R_\odot$ and
$M=20M_\odot$ we find that the specific angular momentum on the stellar 
equator is about $6\times10^{18}\mbox{cm}^2\,\mbox{s}^{-1}$.  On the other
hand the specific angular momentum of marginally bound orbit around
Schwarzschild black hole is
\begin{equation} 
l_{mb}=4\frac{GM_h}{c},
\end{equation} 
For $M_h=20M_\odot$ this gives us only $3.5\times
10^{17}\mbox{cm}^2\,\mbox{s}^{-1}$. Thus, when the star collapses one
would expect the black hole to develop substantial rotation and to 
get surrounded  by an accretion disk towards the final
phase of the collapse \citep{BK09}. The lower accretion rates expected in this
phase make the neutrino mechanism much less effective but do not have much of
effect on the BZ mechanism.  Indeed, assuming the equipartition of gas and
magnetic pressure we find that 
$$
  B^2=8\pi p = 8\pi \alpha^{-2}\rho v_t^2.
$$ 
Applying this at the inner edge ($r=r_g$) of neutrino-cooled accretion disk 
\citep{PWF99} we can estimate the maximum magnetic flux that can be kept on 
the black hole by the disk advection as
\begin{equation} \Psi_h= 3.3\times10^{28}
\!\left(\frac{\alpha}{0.1}\right)^{\!-0.55}
\!\!\left(\frac{M_h}{M_\odot}\right)^{\!1.05}
\!\!\left(\frac{\dot{M\mbox{s}}}{M_\odot}\right)^{\!0.5}
\mbox{G}\,\mbox{cm}^2.
\label{mflux2}
\end{equation} 
For $M_h=20M_\odot$ and $M_\odot=0.01M_\odot/s$ this gives us the formidable 
$\Psi_h\simeq 7\times10^{28}\mbox{G}\,\mbox{cm}^2$. 
According to  eq.\ref{e-bz} much lower fluxes are needed to explain the observed 
energetics of GRBs. 
      
Even faster rotation is expected in the case of merger of helium star with its
compact companion, neutron star or a black hole, during the common envelope
phase of a close binary \citep{TI79,FW98,ZF01}.  Since the orbital angular momentum
of binary star is
\begin{equation} J^2\sub{orb}(L) = \frac{GM^2\sub{h}M^2 L}{M\sub{h}+M}.
\label{b-1}
\end{equation}
where $M\sub{h}$ is the mass of the compact star, we can roughly estimate the
specific orbital angular momentum deposited at the distance $r$ from the
centre of the helium star as
\begin{equation} l^2(r)=J^2\sub{orb}(r)/M^2(r) =
\frac{GM^2\sub{h}r}{M\sub{h}+M(r)},
\label{b-2}
\end{equation}
where $M(r)$ is the helium star mass within radius $r$.  When $M(r)$ is a
slow function, as it is in Bethe's model where $M(r) \propto \ln r$, we may
use in eq.\ref{b-2} the total mass of the helium star, $M$, in place of
$M(r)$.  For $M\sub{bh}=2M_\odot$ and $M\sub{hc}=20M_\odot$ this gives us
\begin{equation} l(r\sub{9})=1.5\times10^{17} \sqrt{r\sub{9}}\,
\mbox{cm}^2\mbox{s}^{-1},
\label{b-3}
\end{equation}
where $r\sub{9}=r/10^9$cm. In spite of all the approximations, such estimates
agree quite well with the results of numerical simulations \citep{ZF01}. For
$r_9>1$ this is higher than the upper value of $l=10^{17}\,
\mbox{cm}^2\mbox{s}^{-1} $ imposed in the collapsar simulations by
\citet{MW99}. The higher angular momentum is one of the factors that reduces
the disk accretion rates and hence the efficiency of the neutrino annihilation
mechanism \citep{ZF01} but, as we have already commented, for the
Blandford-Znajek mechanism this is much less of a problem.

The compact companion may initially be a neutron star but during the in-spiral
it is expected to accumulate substantial mass and become a black hole
\citep{ZF01}. In fact, during the inspiraling the compact companion
accumulates not only mass but spin as well. The final value of the black hole
rotation parameter computed in \citet{ZF01} varies from model to model between
$a=0.14$ to $a=0.985$. Thus, provided that the normal star of the common
envelope binary is a magnetic star we have all the key ingredients for a
successful magnetically driven GRB explosion: (i) a rapidly rotating black
hole, (ii) a rapidly rotating and (iii) compact collapsing star, and (iv) a
``fossil'' magnetic field which is strong enough to tap the black hole energy
at rates required by observations (see eq.\ref{e-bz}).  It has been noted many
times that the magnetic fluxes of magnetic Ap stars, magnetic white dwarfs,
and pulsars are remarkably similar \citep[e.g.][]{FW05}, suggesting that 
they could be related. Now one may consider adding another type of objects to 
this list - GRB collapsars.

\section{Conclusions}

In this paper we continued our study into the potential of the 
Blandford-Znajek mechanism as main driver in the central engines
of Gamma Ray Bursts. In particularly, we analysed the conditions 
for activation of the mechanism in the collapsar model where 
accumulation of magnetic flux by the central black hole is likely 
to be a byproduct of stellar collapse. 

It appears that the rotating black hole begins to pump energy into 
the surrounding and overpowers accretion only when 
the rest mass energy density of matter drops below the energy density of 
the electromagnetic field near the ergosphere (see eq.\ref{cond-1}).  
Only under this condition 
the  MHD waves generated in the ergosphere can transport energy and angular 
momentum away from the black hole. 

In the case of spherical accretion the above criterion can be written as 
a condition on the total magnetic flux accumulated by the black hole 
and the total mass accretion rate (see eq.\ref{cond-2}). However, in the 
case of nonspherical accretion, characteristic for the collapsar model, this 
integral condition becomes less applicable due to the spacial separation 
of magnetic field and accretion flow inside the accretion shock.   
In fact, our numerical simulations show explosions for significantly lower 
magnetic fluxes compared to those expected from the integral criterion.       
We have not included the neutrino heating in our numerical models. 
This is another factor that can make activation of the BZ mechanism 
in the collapsar setup easier.     

Both the general energetic considerations and the condition for 
activation of the Blandford-Znajek mechanism require the central 
black hole to accumulate magnetic flux comparable to the highest 
observed values for magnetic stars, $\Psi\simeq 10^{27}\mbox{G}\,\mbox{cm}^2$.
Current evolutionary models of massive solitary stars indicate that fossil field 
of such strength should slow down the rotation of helium cores below the rate required for formation of accretion disks around black holes of failed supernovae.
The conflict between strong fossil magnetic field and rapid stellar 
rotation required by the BZ mechanism may be resolved in the models 
where the GRB progenitor is a component of close binary. In this case the 
high spin of the progenitor can be sustained at the expense of the 
orbital rotation. Moreover, the strong gravitational interaction between 
components of close binary helps to explain the compactness of GRB progenitors 
deduced from the observed durations of bursts.

The magnetic field can also be generated in the accretion disk and its corona
but in this case the magnetic flux through the black hole ergosphere is not 
sufficiently high to explain the power of hypernovae by the BZ mechanism 
alone. On the other hand, in this case the disk is expected to drive 
MHD wind more powerful than the BZ jet. Although such a wind is unlikely to reach 
the high Lorentz factors deduced from the observations of GRB jets (due to 
high mass loading) it could be behind the observed energetics of hypernovae.


\appendix

\section{Dust accretion in Kerr-Schild coordinates}
\label{a1}
The problem of dust accretion has been considered in \citet{M73}.  In the
coordinate basis of the Boyer-Lindquist coordinates, $\{\partial/\partial
x^{\nu'}\}$ the components of 4-velocity of a dust particle accreting from
rest at infinity and having zero angular momentum are
\begin{equation}
\begin{array}{l} u^{t'}=1+\zeta (r^2+a^2)/\Delta;\\
u^{\phi'}=2a\zeta/\Delta;\\ u^{r'}=-\zeta \eta;\\ u^{\theta'}=0,
\end{array}
\label{u_bl}
\end{equation}
where $\eta=\sqrt{(r^2+a^2)/2r}$.  In order to find the 4-velocity components
in the coordinate basis of Kerr-Schild coordinate one can simply use the
corresponding transformation law \citep{K04a}:
\begin{equation}
\begin{array}{l} dt=dt'+(2r/\Delta)dr';\\ d\phi=d\phi'+(a/\Delta)dr';\\
dr=dr';\\ d\theta = d\theta'.
\end{array}
\label{transf}
\end{equation}
This gives us
\begin{equation}
\begin{array}{l} u^t=1+\zeta\eta/(1+\eta);\\ u^\phi=-a/(A(1+\eta));\\
u^r=-\zeta \eta;\\ u^\theta=0.
\end{array}
\label{u_ks}
\end{equation}
Notice that in both coordinate systems the particles move over a conical
surface, $\theta=$const. Their rotation about the symmetry axis is due to the
inertial frame dragging effect.

In order to find the density distribution consider the steady-state version of
the continuity equation:
$$
  \partial_r (\sqrt{-g}\rho u^r ) = 0,
$$
where $g=-\sin^2\theta A^2$ is the determinant of the metric tensor in the
Kerr-Schild (and Boyer-Lindquist) coordinates.  This shows that
$$
   \sqrt{-g}\rho u^r = C(\theta),
$$
where $C(\theta)$ is some arbitrary function and then that
$$
  \rho(r,\theta) = \frac{C(\theta)}{\sin\theta \sqrt{2r(r^2+a^2)}}.
$$
At infinity this function does not depend on the polar angle only if
$C\propto\sin\theta$ which implies that $\rho$ does not depend on $\theta$ for
any $r$. Denoting as $\rho_+$ the rest mass density at the outer event horizon
we may now write
\begin{equation} \rho=\rho_+ \left(\frac{r_+}{r}\right)\frac{1}{\eta}.
\end{equation}
%

\section{Initial magnetic field}
\label{a2}

In this section we employ the 3+1 splitting of electrodynamics described in 
\citet{K04a} where the electromagnetic field is represented by four 3-vectors,
$\bD,\bE,\bB$, and $\bH$, defined via

\begin{equation} B^i=\alpha \Fs^{it},
\label{B1}
\end{equation}
\begin{equation} E_i =\frac{\alpha}{2} e_{ijk} \Fs^{jk},
\label{E1}
\end{equation}
\begin{equation} D^i=\alpha F^{ti},
\label{D1}
\end{equation}
\begin{equation} H_i =\frac{\alpha}{2} e_{ijk} F^{jk},
\label{H1}
\end{equation} where $\Fs^{\mu\nu}$ is the Faraday tensor, $F^{\mu\nu}$ is the
Maxwell tensor, $e_{ijk}$ is the Levi-Civita tensor of space and $\alpha$ is
the lapse function.  In stationary metric, $\partial_t g_{\nu\mu}=0$, the
evolution of these vector fields is described by the Maxwell equations for
electromagnetic field in matter.  The effects of curved space time are
incorporated via the non-Euclidian spatial metric and the constitutive
equations
\begin{equation} \bE = \alpha \bD + \vpr{\bbeta}{B},
\label{E3}
\end{equation}

\begin{equation} \bH = \alpha \bB - \vpr{\bbeta}{D}.
\label{H3}
\end{equation} where $\alpha$ is the lapse function and $\bbeta$ is the shift
vector.  In terms of these vectors the condition of perfect conductivity,
$F_{\nu\mu}u^\mu=0$, can be written in the familiar form
\begin{equation} \bE=-\vpr{v}{B},
  \label{E}
\end{equation}
where $v^i=dx^i/dt$ is the usual 3-velocity vector.

Both in the Boyer-Lindquist and the Kerr-Schild coordinates the poloidal
component of Poynting flux, $S^i = -\alpha T^i_t$, is given by
\begin{equation} \bS_p=\vpr{E_p}{H_\phi} + \vpr{E_\phi}{H_p}
  \label{S}
\end{equation}
whereas the the poloidal component of the angular momentum flux, $L^i=\alpha
T^i_\phi$, is
\begin{equation} \bL_p=-(\spr{E}{m})\bD_p -(\spr{H}{m})\bB_p,
  \label{L}
\end{equation}
where $\bm=\partial/\partial\phi$.

Suppose that in the Boyer-Lindquist coordinates the magnetic field is purely
radial, $\bB'_\theta=\bB'_\phi=0$.  Since $\bbeta\parallel\bm$ eq.\ref{H3}
then implies that
$$
\bH'_\phi=\alpha\bB'_\phi=0.
$$  
Provided this magnetic field is frozen into the flow described by
eq.\ref{u_bl} we also find that
$$
  \bE'=-\vpr{v'_\phi}{B'_r}.
$$ 
Given these results the equations \ref{L} and \ref{S} ensure that
$$
  \bS'_p=\bL'_p=0.
$$

In order to find the corresponding electromagnetic field in the Kerr-Schild
coordinates we simply apply the transformation law \ref{transf}). This way we
find that
\begin{equation} \bS_p=\bL_p=0
\end{equation} and
\begin{equation} B^\theta=0, \quad B^\phi = B^r
\frac{a(A+2r)}{A\Delta+2r(r^2+a^2)}.
\end{equation} The divergence-free condition requires
$$
  \partial_r(\sqrt{\gamma} B^r)=0
$$ 
and hence
$$
  B^r = \frac{f(\theta)}{\sqrt{\gamma}}
$$
where $\gamma$ is the determinant of the spatial metric. Provided that at
infinity (where $\gamma\propto\sin^2\theta$) $B^r$ does not depend on the
polar angle we then have
\begin{equation} B^r=B_0 \frac{\sin\theta}{\sqrt{\gamma}}.
\end{equation}


\begin{thebibliography}{}
\bibitem[\protect\citeauthoryear{Abt \& Willmarth}{2000}]{AW00}
 Abt H.A., Willmarth D.W.,2000, in eds. K.S.Cheng et al., 
 {\it Stellar Astrophysics}, Kluwer, p.175
\bibitem[\protect\citeauthoryear{Aloy et al.}{2000}]{AIMGM00}
 Aloy M.A., Muller E., Abner J.M., Marti J.M., MacFadyen A.I., 2000, ApJ, 531, L119
\bibitem[\protect\citeauthoryear{Ardeljan et al.}{2005}]{ABM05}
 Ardeljan N.V., Bisnovatyi-Kogan G.S., Moiseenko S.G., 2005, MNRAS, 359, 333
\bibitem[\protect\citeauthoryear{Barkov \& Komissarov}{2008a}]{BK08}
 Barkov M.V., Komissarov S.S, 2008a, MNRAS, 385, L28
\bibitem[\protect\citeauthoryear{Barkov \& Komissarov}{2008b}]{BK08b}
 Barkov M.V., Komissarov S.S, 2008b, in {\it High Energy Gamma-ray Astronomy}, 
 AIP Conference Proceedings, v.1085, p.608
\bibitem[\protect\citeauthoryear{Barkov \& Komissarov}{in preparation}]{BK09}
 Barkov M.V., Komissarov S.S, in prerparation.
\bibitem[\protect\citeauthoryear{Begelman et al.}{1984}]{BBR84}
 Begelman M.C., Blandford R.D., Rees M.J., 1984, Rev.Modern Phys., 56, 255
\bibitem[\protect\citeauthoryear{Bethe}{1990}]{Bethe}
 Bethe H.A., 1990, Rev.Mod.Phys.,62,801
\bibitem[\protect\citeauthoryear{Bezchastnov et~al.}{1997}]{B97}
 Bezchastnov V.G., Haensel P., Kaminker A.D., Yakovlev D.G., 1997, A\&Ap, 328,409
\bibitem[\protect\citeauthoryear{Bisnovatyi-Kogan \& Ruzmaikin}{1974}]{BR74}
 Bisnovatyi-Kogan G.S., Ruzmaikin A.A., 1974, Ap\&SS, 28, 45
\bibitem[\protect\citeauthoryear{Bisnovatyi-Kogan \& Ruzmaikin}{1976}]{BR76}
 Bisnovatyi-Kogan G.S., Ruzmaikin A.A., 1976, Ap\&SS, 42, 401
\bibitem[\protect\citeauthoryear{Blandford \& Znajek}{1977}]{BZ77}
 Blandford R.D. and Znajek R.L., 1977, MNRAS, 179, 433
\bibitem[\protect\citeauthoryear{Blandford \& Payne}{1982}]{BP82}
 Blandford R.D., Payne D.G., 1982, MNRAS, 199, 883
\bibitem[\protect\citeauthoryear{Blondin et al.}{2003}]{BMD03}
 Blondin J.M., Mezzacappa A., DeMarino C., 2003, ApJ, 584,971 
\bibitem[\protect\citeauthoryear{Braithwaite \& Spruit}{2004}]{BS04}
 Braithwaite J., Spruit H.C., 2004, Nature, 431, 819
\bibitem[\protect\citeauthoryear{Buras et al.}{2006}]{BJRK06}
 Buras R., Janka H.-Th., Rampp M., Kifonidis K., 2006, A\&A 457, 281
\bibitem[\protect\citeauthoryear{Burrows et al. }{2007}]{BD07}
 Burrows A., Dessart L., Livne E., Ott C.D., Murphy J., 2007,
 ApJ,664,416
\bibitem[\protect\citeauthoryear{Camenzind}{1989}]{C89}
 Camenzind M., in {\it Accretion disks and magnetic fields in astrophysics},
 ed. G.Belvedere, Kluwer, Dordrecht, p.129, 1989
\bibitem[\protect\citeauthoryear{Donati et al.}{2002}]{D02}
 Donati J.-F., Babel J., Harries T.J., Howarth I.D., Petit P. 
 and Semel M., 2002, MNRAS, 333, 55
\bibitem[\protect\citeauthoryear{Ferrario \& Wickramasinghe}{2005}]{FW05}
 Ferrario L., Wickramasinghe D.T., 2005, MNRAS, 356, 615
\bibitem[\protect\citeauthoryear{Frayer \& Woosley}{1998}]{FW98}
 Frayer C.L, Woosley S.E., 1998, ApJ Lett., 502, L9
\bibitem[\protect\citeauthoryear{Fujimoto et al.}{2006}]{FKYHS06}
 Fujimoto S., Kotake K., Yamada S., Hashimoto M., Sato K., 2006, ApJ, 644, 1040
\bibitem[\protect\citeauthoryear{Ghosh \& Abramowicz}{1997}]{GA97}
 Ghosh P., Abramowicz M.A., 1997, MNRAS, 292, 887
\bibitem[\protect\citeauthoryear{Heger et al.}{2000}]{HLW00}
 Heger A., Langer N., Woosley S.E., 2000, ApJ, 528, 368
\bibitem[\protect\citeauthoryear{Heger et al.}{2005}]{HWS05}
 Heger A., Woosley S.E., Spruit H.C., 2005, ApJ, 626, 350
\bibitem[\protect\citeauthoryear{Meszaros \& Rees}{1997}]{MR97}
 M\'esz\'aros P., Rees M.J., 1997, ApJ Lett., 482, L29 
\bibitem[\protect\citeauthoryear{Hirschi et al.}{2004}]{HMM04}
 Hirschi R., Meynet G., Maeder A, 2004, A\&A, 425, 649
\bibitem[\protect\citeauthoryear{Hirschi et al.}{2005}]{HMM05}
 Hirschi R., Meynet G., Maeder A, 2005, A\&A, 443, 581
\bibitem[\protect\citeauthoryear{Heyvaerts et al.}{1996}]{HPB96}
 Heyvaerts J., Priest E.R., Bardou A., 1996, ApJ, 473, 403
\bibitem[\protect\citeauthoryear{Howarth et al.}{1997}]{H97}
 Howarth I.D., Siebert K.W., Hussain G.A., Prinja R.K, 1997, MNRAS, 284, 265
\bibitem[\protect\citeauthoryear{Igumenshchev}{2008}]{I08}
 Igumenshchev I.V., 2008, ApJ, 677, 317
\bibitem[\protect\citeauthoryear{Ivanova et al.}{1969}]{IIN69}
 Ivanova L.N., Imshennik V.S., Nadezhin D.K., 1969, 
 Sci.Inf.Astr.Council.Acad.Sci, 13,3
\bibitem[\protect\citeauthoryear{Koide}{2004}]{Koi04}
 Koide S.,2004,ApJ Lett., 606,L45
\bibitem[\protect\citeauthoryear{Komissarov}{1999}]{K99}
 Komissarov S.S., 1999, MNRAS, 303, 343
\bibitem[\protect\citeauthoryear{Komissarov}{2001}]{K01}
 Komissarov S.S., 2001,MNRAS,326,L41
\bibitem[\protect\citeauthoryear{Komissarov}{2004a}]{K04a}
 Komissarov S.S., 2004a,MNRAS,350,427
\bibitem[\protect\citeauthoryear{Komissarov}{2004b}]{K04b} 
 Komissarov S.S., 2004b, MNRAS, 350, 1431
\bibitem[\protect\citeauthoryear{Komissarov}{2006}]{K06} 
 Komissarov S.S., 2006, MNRAS, 368, 993
\bibitem[\protect\citeauthoryear{Komissarov}{2008}]{K08} 
 Komissarov S.S., 2008, arXiv0804.1912
\bibitem[\protect\citeauthoryear{Lee et al.}{2000}]{LBW00}
 Lee H.K., Brown G.E., Wijers R.A.M.J., 2000, ApJ, 536, 416 
\bibitem[\protect\citeauthoryear{Livio et al.}{1999}]{LOP99} 
 Livio M., Ogilvie G.I., Pringle J.E., 1999, ApJ, 512, 100
\bibitem[\protect\citeauthoryear{Lubow et al.}{1994}]{LPP94} 
 Lubow S.H., Papaloizou J.C.B., Pringle J.E., 1994, MNRAS, 267, 235
\bibitem[\protect\citeauthoryear{Macdonald \& Thorne}{1982}]{MT82} 
 Macdonald D.A. and Thorne K.S., 1982, MNRAS, {\bf 198}, 345.
\bibitem[\protect\citeauthoryear{MacFadyen \& Woosley}{1999}]{MW99} 
 MacFadyen A.I. \& Woosley S.E, 1999, ApJ, 524, 262
\bibitem[\protect\citeauthoryear{Maeder \& Meynet}{2005}]{MM05}
 Maeder A., Meynet G., 2005, A\&A, 440, 1041
\bibitem[\protect\citeauthoryear{McKinney \& Gammie}{2004}]{MG04}
 McKinney J.C., Gammie C.F., 2004, ApJ, 611, 977
\bibitem[\protect\citeauthoryear{McKinney}{2006}]{M06}
 McKinney J.C., 2006, MNRAS,368,1561
\bibitem[\protect\citeauthoryear{Mignone \& Bodo}{2006}]{MB06}
 Mignone A., Bodo G., 2006, MNRAS, 368, 1040
\bibitem[\protect\citeauthoryear{Misner et al.}{1973}]{M73}
 Misner C.W., Thorne K.S., Wheeler J.A., 
 {\it Gravitation}, San Francisco: W.H. Freeman and Co., 1973
\bibitem[\protect\citeauthoryear{Mizuno et al.}{2004a}]{MYKS04a}
 Mizuno Y., Yamada S., Koide S., Shibata K., 2004a, ApJ, 606, 395
\bibitem[\protect\citeauthoryear{Mizuno et al.}{2004b}]{MYKS04b}
 Mizuno Y., Yamada S., Koide S., Shibata K., 2004b, ApJ, 615, 389
\bibitem[\protect\citeauthoryear{Moiseenko et al.}{2006}]{MBA06}
 Moiseenko S.G., Bisnovatyi-Kogan G.S., Ardeljan N.V., 2006, MNRAS, 370, 501
\bibitem[\protect\citeauthoryear{Moss}{1987}]{M87}
 Moss D.,1987, MNRAS, 226, 297
\bibitem[\protect\citeauthoryear{Nagakura \& Yamada}{2009}]{NY09}
 Nagakura H., Yamada S., 2009, arXiv:0901.4053
\bibitem[\protect\citeauthoryear{Nagataki et al.}{2007}]{NTMT07}
 Nagataki S., Takahashi R., Mizuta A., Takiwaki T., 2007, ApJ, 659, 512
\bibitem[\protect\citeauthoryear{Nagataki}{2009}]{N09}
 Nagataki S., 2009, arXiv:0902.1908 
\bibitem[\protect\citeauthoryear{Narayan et al.}{1992}]{NPP92} 
 Narayan R., Paczy\'nski B., Piran T., 1992, ApJ Lett., 395, L8
\bibitem[\protect\citeauthoryear{Obergaulinger et al.}{2006}]{OAM06}
 Obergaulinger M., Aloy M.A., Dimmelmeier H., Muller E.,
 2006,A\&A,457,209
\bibitem[\protect\citeauthoryear{Penny}{1996}]{P96}
 Penny L.R., 1996, ApJ., 463, 737
\bibitem[\protect\citeauthoryear{Piran}{2005}]{P05}
 Piran T., 2005, Rev.Mod.Phys., 76, 1143
\bibitem[\protect\citeauthoryear{Popham et al.}{1999}]{PWF99}
 Popham R., Woosley S.E. and Fryer C.L., 1999, ApJ, 518, 356
\bibitem[\protect\citeauthoryear{Proga et al.}{2003}]{PMAB03}
 Proga D., MacFadyen A.I., Armitage P.J., Begelman M.C., 2003, ApJ, 629, 397
\bibitem[\protect\citeauthoryear{R\"adler}{1968}]{R68}
 R\"adler K.-H., 1968, Z.Naturforsch., A, 23, 1851
\bibitem[\protect\citeauthoryear{Robstein \& Lovelace}{2008}]{RL08}
 Robstein D.M., Lovelace R.V.E., 2008, ApJ, 677, 1221
\bibitem[\protect\citeauthoryear{Schinder et al.}{1987}]{S87}
 Schinder P.J., Schramm D.N., Wiita P.J., Margolis S.H., Tubbs D.L., 1987, ApJ, 313, 531.
\bibitem[\protect\citeauthoryear{Sekiguchi \& Shibata}{2007}]{SS07}
 Sekiguchi Y., Shibata M.,2007,Prog.Theor.Phys,117,1029
\bibitem[\protect\citeauthoryear{Scheck et al.}{2008}]{SJFK08}
 Scheck L., Janka H.-Th., Foglizzio T., Kifonidis, K., 2008, A\&A, 477, 931
\bibitem[\protect\citeauthoryear{Schmidt et al.}{2003}]{S03}
 Schmidt et al., 2003, ApJ, 595, 1101
\bibitem[\protect\citeauthoryear{Spruit}{2002}]{S02}
 Spruit H.C., 2002, A\&A, 381, 923
\bibitem[\protect\citeauthoryear{Spruit \& Uzdensky}{2005}]{SU05}
 Spruit H.C., Uzdensky D.A., 2002, ApJ, 629, 960
\bibitem[\protect\citeauthoryear{Takahashi et al.}{1990}]{T90}
 Takahashi M., Niita S., Tatematsu Y., Tomimatsu A., 1990, ApJ, 363, 206.
\bibitem[\protect\citeauthoryear{Tao et al.}{1998}]{T98}
 Tao L., Proctor M.R.E., Weiss N.O., 1998, MNRAS, 300, 907
\bibitem[\protect\citeauthoryear{Taam \& Sandquist}{2000}]{TaS00}
 Taam R.E., Sandquist E.L., 2000, Ann.Rev.A\&A, 38, 113
\bibitem[\protect\citeauthoryear{Tayler}{1973}]{T73} 
 Tayler R., 1973, MNRAS, 161, 365
\bibitem[\protect\citeauthoryear{Timmes \& Swesty}{2000}]{TS00} 
 Timmes F.X., Swesty F.D., 2000, ApJSS, 126, 501
\bibitem[\protect\citeauthoryear{Tout \& Pringle}{1996}]{TP96} 
 Tout C.A., Pringle J.E., 1996, MNRAS, 281, 219
\bibitem[\protect\citeauthoryear{Tutukov \& Iungelson}{1979}]{TI79}
 Tutukov A., Iungelson L., in {\it Mass loss and evolution of O-type stars}, 
 Redel, Dordrecht, p.401, 1979
\bibitem[\protect\citeauthoryear{Uzdensky \& MacFadyen}{2006}]{UM06} 
 Uzdensky D.A. \& MacFadyen A.I., 2006, ApJ, 647, 1192
\bibitem[\protect\citeauthoryear{Woosley}{1993}]{W93}
 Woosley S.E., 1993, ApJ, 405, 273
\bibitem[\protect\citeauthoryear{Woosley \& Baron}{1992}]{WB92} 
 Woosley S.E., Baron E., 1992, ApJ, 391, 228
\bibitem[\protect\citeauthoryear{Woosley \& Bloom}{2006}]{WB06} 
 Woosley S.E., Bloom J.S., 2006, Ann.Rew.A\&A, 44, 507
\bibitem[\protect\citeauthoryear{Woosley \& Heger}{2006}]{WH06} 
 Woosley S.E., Heger A., 2006, ApJ, 637, 914
\bibitem[\protect\citeauthoryear{Yoon et al.}{2006}]{YCL06} 
 Yoon S.-C., Langer N., Norman C., 2006, A\&A, 460, 199.
\bibitem[\protect\citeauthoryear{Yoon et al.}{2008}]{YCL08} 
 Yoon S.-C., Cantiello M., Langer N.,2008, arXiv:0801.4373
\bibitem[\protect\citeauthoryear{Zel'dovich}{1957}]{Z57} 
 Zel'dovich Ya.B., 1957, Sov.Phys.JETP, 4, 460
\bibitem[\protect\citeauthoryear{Zhang \& Fryer}{2001}]{ZF01}
 Zhang W., Fryer C.L., 2001, ApJ, 550, 357
 
\end{thebibliography}
\end{document}